\pdfoutput=1
\documentclass[11pt,a4paper]{article}
\usepackage[utf8]{inputenc}
\usepackage{a4wide}
\usepackage{amsmath,amssymb,graphicx}
\usepackage[caption=false]{subfig}
\usepackage{bm,dsfont}
\usepackage[colorlinks=true,allcolors=blue]{hyperref}
\usepackage[square,numbers,sort&compress]{natbib}

\newcommand{\beq}{\begin{eqnarray}}
\newcommand{\eeq}{\end{eqnarray}}
\newcommand{\non}{\nonumber\\}

\newcommand{\p}{\partial}
\newcommand{\Tr}{\qopname\relax o{Tr}}

\newcommand{\SU}{\mathrm{SU}}
\newcommand{\U}{\mathrm{U}}
\renewcommand{\d}{{\mathrm{d}}}
\renewcommand{\i}{\mathrm{i}}

\newcommand{\dA}{\delta{\mkern-2.5mu}A}
\newcommand{\dmu}{\delta{\mkern-1mu}\mu}

\newcommand{\bchi}{\boldsymbol{\chi}}
\newcommand{\btau}{\boldsymbol{\tau}}

\newcommand{\calA}{\mathcal{A}}
\newcommand{\calF}{\mathcal{F}}
\newcommand{\calL}{\mathcal{L}}
\newcommand{\calO}{\mathcal{O}}
\newcommand{\calR}{\mathcal{R}}
\newcommand{\wha}{\widehat{a}}
\DeclareMathOperator{\MeV}{MeV}
\DeclareMathOperator{\fm}{fm}
\newcommand{\sfL}{\mathsf{L}}

\newcommand{\AdS}[1]{AdS$_{#1}$}

\begin{document}
\pagenumbering{Roman}
\begin{titlepage}
  \phantom{.}
  \vskip2cm
  \begin{center}
    {\Large\bf Symmetry energy in holographic QCD}\\[2cm]
    {\bf Lorenzo Bartolini, Sven Bjarke Gudnason}\\[2cm]
    {Institute of Contemporary Mathematics, School of Mathematics and Statistics,\\ Henan University, Kaifeng, Henan 475004, P.~R.~China}
  \end{center}
  \vfill
  \begin{abstract}
  We study the symmetry energy (SE), an important quantity in nuclear
  physics, in the Witten-Sakai-Sugimoto model and in a much simpler
  hard-wall model of holographic QCD. The SE is the energy
  contribution to the nucleus due to having an unequal number of
  neutrons and protons. Using a homogeneous Ansatz representing
  smeared instantons and quantizing their isospin, we extract the SE
  and the proton fraction assuming charge neutrality and
  beta-equilibrium, using quantization of the isospin zeromode. We
  also show the equivalence between our method adapted from solitons
  and the usual way of the isospin controlled by a chemical potential at
  the holographic boundary.
  We find that the SE can be well described in the WSS 
  model if we allow for a larger 't Hooft coupling and lower
  Kaluza-Klein scale than is normally
  used in phenomenological fits.
  \end{abstract}
  \vfill
  \rule{10cm}{0.5pt}\\
  {\tt lorenzo@henu.edu.cn}, {\tt gudnason@henu.edu.cn}
\end{titlepage}

\pagenumbering{arabic}

\tableofcontents

\section{Introduction}

The equation of state (EOS) of nuclear matter is central to nuclear
physics from neutron stars to heavy ion collisions, and an important
feature is the symmetry energy (SE) as a function of the density.
The symmetry energy is the symmetric increase in energy as one moves
away from the isospin symmetric point, that is, the point where the
number of protons equals that of neutrons, i.e.~
$E(\rho)=E_0(\rho)+S(\rho)\beta^2+\cdots$, with $\beta=\frac{N-Z}{A}$
being the difference between the number of neutrons $N$ and the number of
protons $Z$, normalized by the atomic mass number, $A=Z+N$ ---
for a nice review see Ref.~\cite{Baldo:2016jhp}.
The symmetry energy is experimentally well constrained around
saturation density, $\rho_0\sim 0.16$fm$^{-3}$, to be near
$S(\rho_0)\sim 30$ MeV -- both from astrophysical
observations as well as heavy ion collision data -- but much less so
at larger densities. 
The symmetry energy around saturation density is conventionally
expanded as
\beq
S(\rho) = S_0 + \frac13L\epsilon + \frac{1}{18}K_{\rm sym}\epsilon^2 + \cdots
\eeq
with $\epsilon:=(\rho-\rho_0)/\rho_0$, whereas $L$ and $K_{\rm sym}$
are proportional to the slope and second
derivative of the SE with respect to the density.
Expectedly, the constraints on $L$ and $K_{\rm sym}$ are less tight
than those on $S_0$.
Traditionally, the symmetry energy was defined for nuclear matter,
which can be thought of as an infinitely large nucleus at density
$\rho$, and so surface effects are absent.
The symmetry energy can equally well be defined for a fixed, but
finite, atomic number $A$.

Current experimental bounds on the first 3 observables of the symmetry
energy as in the expansion of the density come from mass, radius and
tidal deformation of neutron stars, excitation energies of isobaric
analog states, neutron skin in Sn isotopes and $^{208}$Pb as well
as heavy ion collision data
\cite{Danielewicz:2013upa,FiorellaBurgio:2018dga,Li:2021thg,Tang:2021snt,Reed:2021nqk,Tang:2021snt,Essick:2021ezp,Gil:2021ols,deTovar:2021sjo}.

The equation of state in nuclear physics relates the energy density
with the pressure and is the main ingredient in the understanding of
neutron stars as well as heavy ion collisions.
The problem with obtaining the equation of state for nuclei is that
the strong nuclear force is governed by Quantum Chromodynamics (QCD),
an inherently strongly coupled theory and hence cannot be tackled by
perturbation theory or first-principles calculations.
Nuclear physics, in particular, ab initio methods, like the no-core
shell model \cite{Barrett:2013nh}, utilize pion scattering data to reconstruct the
interaction potential of nuclei and this approach leads to solid
predictions for the interaction potential and the chiral effective field theory can accurately determine the EOS \cite{Tews:2012fj}, albeit only at relatively small densities.
QCD at high energies is perturbative due to its asymptotically free
nature, and hence can be used to make solid predictions for the EOS
\cite{Kurkela:2009gj}, unfortunately at pressures far larger than
those of a neutron star -- the most compact object known, not
collapsed into a black hole (BH).

A new paradigm of studying QCD and attempting to extract observables
for nuclear physics and hadronic physics, was envisioned by Maldacena
at the end of the '90-ies \cite{Maldacena:1997re} and further
elaborated by Witten \cite{Witten:1998zw}.
After a couple of decades, the mentioned framework known as holography
or AdS/CFT, has been coined holographic QCD (HQCD) when applied to the
strong nuclear force \cite{Kim:2012ey,Rebhan:2014rxa,Ammon:2015wua}.
There are two main approaches to HQCD, top-down and bottom-up; the
top-down approach is based on string-theory constructions and the most
prominent example is the Witten-Sakai-Sugimoto model (WSS)
\cite{Witten:1998zw,Sakai:2004cn,Sakai:2005yt}.
For the bottom-up construction, which shares similar theoretical
ingredients, two main types of models are known as soft-wall (SW)
(e.g.~Improved HQCD \cite{Gursoy:2007cb,Gursoy:2007er}
and V-QCD \cite{Jarvinen:2011qe,Alho:2013hsa,Jarvinen:2021jbd}) and 
hard-wall (HW) models \cite{Polchinski:2001tt,Boschi-Filho:2002wdj,Polchinski:2002jw,Boschi-Filho:2002xih,deTeramond:2005su,Erlich:2005qh,DaRold:2005vr,Hirn:2005nr,Karch:2006pv}.
Especially, the top-down type of HQCD have quite some predictive
power, in the sense that the models have very few adjustable
parameters \cite{Kim:2012ey,Rebhan:2014rxa,Ammon:2015wua}.
For the WSS, there is the mass scale and the 't Hooft coupling, where
the mass scale is normally fitted to the mass of the $\rho$ meson
and the 't Hooft coupling is determined from the pion decay constant
\cite{Sakai:2004cn}.

Attempts have already been made at extracting the SE from various HQCD
models, including top-down approaches as in the D4/D6 model \cite{Kim:2010dp} and even the WSS model, see Ref.~\cite{Kovensky:2021ddl}, and
NSs have been constructed using the holographically extracted EOS to
solve the governing Tolman-Oppenheimer-Volkov (TOV) equations
\cite{Kovensky:2021kzl,Bartolini:2022rkl}.
The SE, however, comes out too large in the WSS
\cite{Kovensky:2021ddl}.
In HQCD in contrast to traditional nuclear physics, the proton and the
neutron are not point particles, but are described by a topological
soliton, the Sakai-Sugimoto soliton
\cite{Sakai:2004cn,Hata:2007mb,Hashimoto:2008zw,Bolognesi:2013nja},
which is initially isospin symmetric -- that is, the proton is equal
to the neutron.
In order to compute the SE, we must distinguish the proton from the
neutron and this can be done by the introduction of an isospin
chemical potential \cite{Kovensky:2021ddl}.

HQCD at finite isospin chemical potential has been
object of inspection in the context of many models: top-down approaches include the
D3/D7 model \cite{Erdmenger:2007ap,Erdmenger:2007ja,Erdmenger:2008yj} 
and the WSS model \cite{Parnachev:2007bc,Aharony:2007uu} in the case
of the DBI action and without employing the homogeneous Ansatz (see
below).

In this paper, we propose using the homogeneous Ansatz in the WSS,
but quantizing the isospin symmetry -- a technique known from the
Skyrme model \cite{Adkins:1983ya,Lee:2010sw,Adam:2022aes}, which is the
leading-order low-energy effective theory of the WSS model
\cite{Sakai:2004cn}.
The homogeneous Ansatz represents an approximation to describe 
densely packed nucleons that form nuclear matter above saturation density.
It relies on the assumption that nuclear matter forms a spatially homogeneous
distribution, in which nucleons lose their individual properties: 
despite being shown in Ref.~\cite{Rozali:2007rx} that such a 
configuration is not admitted in holographic models under assumptions
of regularity of the gauge fields, the Ansatz can still be employed with 
modifications, such as formulating it at the level of the field strengths
\cite{Elliot-Ripley:2016uwb} or (as we will do) introducing a discontinuity that acts as
a source of baryon number \cite{Li:2015uea}.
The quantization of the isospin symmetry introduces the isospin
quantum number, which makes it possible to extract the SE as the
coefficient of the square of the difference between the number of
protons and neutrons.
The quantization method of introducing isospin is also shown to be
equivalent to using a chemical potential, see App.~\ref{A:mu}.
We find a lower SE compared to previous attempts in the WSS
\cite{Kovensky:2021ddl}, since we include all the needed fields in our
Ansatz and because we choose a different $N_c$ scaling such that the
nucleons are states with the minimal isospin quantum number; however, there is no
difference coming from using either the chemical potential or the
quantization method -- they are equivalent as shown explicitly in
App.~\ref{A:mu}.
In particular, we find a phenomenologically viable value of the constant
$S(\rho_0)$ at saturation density and the first two coefficients $L$
and $K_{\rm sym}$ are compatible with current experimental bounds from
astrophysics and heavy ion collision data for a certain choice of the model
parameters.

\section{Model}

We will treat the WSS and the HW model on equal footing in the
following.
The model at low energies is described by the Yang-Mills (YM) and
Chern-Simons (CS) actions in 5-dimensional \AdS{5} (or \AdS{5}-like, for the WSS model) spacetime $(M,g)$:
\begin{align}
  S_{\rm YM} &= -\kappa\Tr\int_M \calF\wedge*\calF,\non
  S_{\rm CS} &= \frac{N_c}{24\pi^2}\Tr\int_M
  \left[\calA\wedge\calF^2
  -\frac{\i}{2}\calA^3\wedge\calF
  +\frac{1}{10}\calA^5\right],
  \label{eq:action}
\end{align}
with $S=S_{\rm YM}+S_{\rm CS}$ being the total action, the constant
$\kappa=\frac{\lambda N_c}{216\pi^3}$ for the WSS model and
$\kappa=M_5$ for the HW model, $\lambda=g_{\rm YM}^2 N_c$ the 't Hooft
coupling, $N_c$ the number of colors of QCD (i.e.~3 in nature), the
field strength 2-form is
\beq
\calF=\tfrac12(\p_\alpha\calA_\beta-\p_\beta\calA_\alpha+\i[\calA_\alpha,\calA_\beta])\d x^\alpha\wedge\d x^\beta,
\eeq
$\alpha,\beta=0,1,2,3,4$ with $x^4=z$ the holographic coordinate, 
and the power of forms is understood with the wedge product.
The metric is
\beq
g = h(z)k(z) \d x_\mu \d x^\mu + h^2(z) \d z^2,
\eeq
with $h(z)=k^{-1/3}(z)=(1+z^2)^{-1/3}$, $z\in(-\infty,\infty)$ for WSS and
$h(z)=k(z)=\sfL/z$, $z\in[0,\sfL]$ for the HW model, the index
$\mu=0,1,2,3$ is summed over in the metric and $\mu$ is raised with
the Minkowski metric here.
It is convenient to work in dimensionless units. Both models have a free mass scale
$M_{\rm KK}$ (WSS) and $\sfL^{-1}$ (HW): the WSS model is already presented
with the choice $M_{\rm KK}=1$, to which we add the choice $\sfL=1$ for the HW model. 
The correct powers of the energy scales $M_{\rm KK}$, $\sfL^{-1}$
can then easily be restored via dimensional analysis.

The gauge field can be decomposed for later convenience in the
Abelian and non-Abelian parts as
\beq
\mathcal{A}_{\alpha}={A}_{\alpha }^a T^a + \widehat{A}_{\alpha}\frac{\mathds{1}}{2}, 
\eeq
where the generators of $\SU(2)$ $T^a$ are chosen as $T^a = \frac{1}{2}\tau^a$ so that $\Tr T^aT^b=\frac{1}{2}\delta^{ab}$, and the spacetime indices follow the convention:
\beq
\alpha,\beta,\ldots = \{0,M\} ;\quad M,N,\ldots = \{i,z\} ;\quad \mu,\nu,\ldots = \{0,i\}.
\eeq 
In writing Eq.~\eqref{eq:action}, we performed dimensional reduction
in the WSS, integrating out $S^4$ from the original nine-dimensional
action for the stack of $D8-$Branes, while in the HW we do not
explicitly include an action for the scalar field encoding chiral
symmetry breaking, since we set the scalar field to zero, which is
appropriate in the homogeneous baryonic phase, following
Ref.~\cite{Bartolini:2022rkl}.
Despite not appearing explicitly in our computation, the scalar field
plays an important role: its vacuum energy, determines the density of nuclear matter at the baryonic
onset, hence defining saturation density within this model. For
details on how the scalar field defines the saturation density, but
otherwise vanishes in the baryonic phase, see
App.~\ref{app:scalardecoupling}.
Here we will utilize the fit found in Ref.~\cite{Bartolini:2022rkl}
and only adjust the overall energy scale.

Two further steps were employed in order to write the action and
equations of motion for the two models in a compact way.
For the HW model we assumed the symmetry properties for the fundamental
fields $\mathcal{L}_M,\mathcal{R}_M$ as follows: 
\begin{align}
	\mathcal{L}_i &= -\mathcal{R}_i ,\qquad
	\mathcal{L}_0 =	\mathcal{R}_0 .
\end{align}
For the WSS model, we similarly assumed parity properties of the fields with respect to $z$:
\begin{align}
	\mathcal{A}_i(z) &= -	\mathcal{A}_i(-z), \qquad
	\mathcal{A}_0(z) =	+	\mathcal{A}_0(-z). \label{eq:WSSparity}
\end{align}
With these procedures, we halve the number of fields in the HW model (from
$\mathcal{L},\mathcal{R}$ to $\mathcal{A}$) and the integration
interval in the WSS model (from $(-\infty,+\infty)$ to $[0,+\infty)$)
generating an overall factor of $2$ in the action in both cases.

As a last step, we introduce generic symbols $z_{\rm IR},z_{\rm UV}$
to indicate the infrared and ultraviolet boundary values of the
holographic coordinate\footnote{In the WSS model, the spatial manifold
only has a UV boundary at $z=\pm \infty$. However, when we introduced
the "folding" of the coordinate $z$ exploiting the assumptions
\eqref{eq:WSSparity}, we effectively introduced an IR boundary at the
folding point $z_{\rm IR}=0$. Moreover, the homogeneous Ansatz will
introduce a discontinuity in the field $A_i$ at that point.}, which in
the two models assume the values
\beq
z_{\rm IR} = \begin{cases}
 0 & \text{WSS}\\
 1 & \text{HW}
\end{cases}\qquad , \qquad z_{\rm UV} = \begin{cases}
+\infty & \text{WSS}\\
0 & \text{HW}
\end{cases}
\eeq 
The classical homogeneous Ansatz for isospin-symmetric matter,
reasonable for large-density computations, is defined as
\begin{align}
  \calA^{\rm cl}_0 = \tfrac12\wha_0,\quad
  \calA^{\rm cl}_i = -\tfrac12H\tau^i,\quad
  \calA^{\rm cl}_z = 0,
  \label{eq:Ansatz}
\end{align}
where $\{\wha_0,H\}=\{\wha_0,H\}(z)$ are functions
of the holographic coordinate $z$. We have suppressed the unit 2-by-2 matrices in the terms without a
Pauli matrix $\btau$.

The function $H(z)$ encodes the baryonic density through its value at
$z=z_{\rm IR}$: if either $H(z_{\rm IR})$ or $H'(z_{\rm IR})$ vanish,
then the baryon number would also vanish as noted in
Ref.~\cite{Rozali:2007rx}, so we will assume that 
$H(z)$ obeys a Dirichlet boundary condition $H(z_{\rm IR})=H_0$, with
the value of $H_0$ to be determined by minimization of the
action. This defines the baryon density $\rho$ (assuming
$H(z)\rightarrow0$ for $z\rightarrow z_{\rm UV}$) as follows:
\begin{align}
\rho&=\frac{1}{16\pi^2}\int\d z\;\epsilon^{MNPQ}\Tr F_{MN}F_{PQ}\nonumber\\
&=-\frac{3}{4\pi^2}\int\d z\; H'H^2 \nonumber \\
&=-\epsilon\frac{1}{4\pi^2}\left[H^3\right]_{z_{\rm IR}}^{z_{\rm UV}},
\end{align}
so that the infrared boundary condition for the numerical integration
of the function $H(z)$ is directly related to the baryon number
density as: 
\beq
H(z_{IR}) = \epsilon\left(4\pi^2\rho\right)^\frac{1}{3},
\eeq
where for convenience of putting the two models on same footing, we have
defined the integral in the holographic direction as
\beq
\int\d z\;f(z) := \epsilon\int_{z_{\rm IR}}^{z_{\rm UV}} \d z\;f(z),
\eeq
which we will use throughout the paper and $\epsilon$ assumes
a different sign depending on the model:
\beq
\epsilon = \begin{cases}
+1 & \text{WSS}\\
-1 & \text{HW}
\end{cases}
\eeq
Thus the integral is defined in such a way to take into account the
different orientation in the integration along $z$, dictated by the
choice of coordinates for the two models.
Note that this choice of boundary condition for $H(z_{\rm IR})$ means
that in the WSS, once we restore the full domain of integration
$z\in(-\infty,\infty)$, the function $H(z)$ will be
discontinuous. This still leads to a continuous field strength, since
both $H'$ and $H^2$ are continuous functions. For the HW model
instead, this choice just means that we cannot enforce the standard
boundary condition $L_\mu(z_{\rm IR})-R_\mu(z_{\rm IR})=0$, which has
to be replaced with the one above, implying
$L_\mu(z_{\rm IR})=-R_\mu(z_{\rm IR})$.

\section{Time-dependent configurations}
 
We wish to include the effects of isospin asymmetry in the system. To
do so, we follow a method inspired by the single-soliton analysis: we
know that for the single baryon, the proton and the neutron are described as
degenerate (in absence of quark mass terms\footnote{See
Ref.~\cite{Bigazzi:2018cpg} for the effect of
breaking the degeneracy for the WSS model, when including the quark
mass terms.
For some recent results regarding nuclear matter that include quark
masses in the WSS model, see
Refs.~\cite{Kovensky:2019bih,Kovensky:2020xif,Kovensky:2023mye}.
For results on the phase diagram of a HW model including the effects of quark mass,
see Ref.~\cite{Singh:2022obu}.}) quantum states of the
effective Hamiltonian obtained by considering a slow rotation in
$\SU(2)$. The homogeneous Ansatz shares a similar structure with the
single-soliton configuration, made easier by the absence of
translational moduli\footnote{The translational
moduli $X_i$ are absent because of the assumption of homogeneity,
while the pseudo-modulus size $\rho$ is fixed by the numerical
solution so as to minimize energy. The pseudo-modulus $Z$ describing the
center of the soliton in $z$ is fixed by our Ansatz to be at the
position of the discontinuity. This in principle can also be
determined by choosing $Z=z_0$ that minimizes the free energy as
opposed to our simpler choice $z_0=0$ for all densities. See
Ref.~\cite{CruzRojas:2023ugm} for the inclusion of this effect in the
static approximation.} $X_M$ and $\rho$ (but with the minor complication of not having
an analytical configuration to approximate our static Ansatz
\eqref{eq:Ansatz}), so we can attempt to follow steps similar to the ones
in Refs.~\cite{Hata:2007mb} and \cite{Hashimoto:2008zw}, in order to
obtain a time-dependent configuration -- yet to be quantized.

We start by assuming a configuration of the form 
\begin{align}
	A_0 &= 0,\\
	A_i &= V A_i^{\rm cl} V^{-1} - \i V \partial_i V^{-1},\\
	A_z &= -\i V \partial_z V^{-1},
\end{align}
which implies the following transformations in the field strengths:
\begin{align}
	F_{MN} &= V F^{\rm cl}_{MN} V^{-1}, \\
	F_{0z} &= -VD_z^{\rm cl} \Phi V^{-1}, \\
	F_{0i} &= 0,
\end{align}
where $V(z,t)$ encodes the time-dependent rotation in $\SU(2)$, and $\Phi$ is defined as 
\beq
\Phi \equiv -\i V^{-1}\dot{V}.\label{eq:Phidef}
\eeq
Notice, this is not a gauge transformation since the field $A_0$ is
not transformed along with the rest.
The function $V(z,t)$ needs to depend on $z$ in order to allow us to satisfy the equation of motion
\beq
-\kappa \left(h(z)D_jF^{0j}+D_z\left(k(z)F^{0z}\right)\right)+\frac{N_c}{64\pi^2}\epsilon^{0\alpha_1\alpha_2\alpha_3\alpha_4}F_{\alpha_1\alpha_2}\widehat{F}_{\alpha_3\alpha_4}=0.\label{eq:EOMA0}
\eeq

The function $V(z,t)$ is holographically dual to the $\SU(2)$-valued collective coordinate $a(t)$, as we choose it such that
\beq
V(z\rightarrow z_{\rm UV},t) = a(t),
\eeq
which in turn implies 
\beq
\Phi (z\rightarrow z_{\rm UV},t) = -\i a^{-1}\dot{a} \equiv \frac12 \bchi\cdot\btau, \label{eq:PhiChi}
\eeq
where $\bchi$ is the boundary angular velocity.
The presence of a nonvanishing $F_{0z}$ will also enable a source term
for the fields $\widehat{A}_i$ via the Chern-Simons action, so we will
have to complete the field content by turning on
$\widehat{A}_i = -\frac12 L\chi^i$: here we already guessed that the vector field will be
proportional to the angular velocity $\chi^i$, and we can do so
without loss of generality, since in the homogeneous case this is the
only three-vector available to the Abelian field.

At this stage the problem is well posed and the function $\Phi(z,t)$
can be found by solving Eq.~\eqref{eq:EOMA0}, but it is more
convenient to perform a gauge transformation to make the system easier
to treat.

We perform the gauge $\SU(2)$ transformation 
\beq
A_\alpha \rightarrow A^S_{\alpha}= G A_\alpha G^{-1} -\i G\partial_{\alpha}G^{-1}, \qquad G\equiv aV^{-1},
\qquad \alpha = 0,1,2,3,4,
\eeq
where the superscript ``$S$'' stands for ``singular'', because this is
reminiscent of the transformation changing from the regular gauge to
the singular gauge in the single-soliton case.
With this choice (dropping the superscript ``$S$'' for convenience,
since we will use this gauge henceforth) the field content becomes 
\begin{align}
  A_0 &= a\left(\Phi-\frac12 \btau\cdot\bchi\right)a^{-1},\\
  A_i &= a A_i^{\rm cl} a^{-1},\\
  A_z &= 0.
\end{align}
Now we can factorize the function $\Phi(z,t)$ as 
\beq
\Phi = \Phi^a\chi^a \equiv \widetilde{G} \bchi\cdot\btau,
\eeq
and since we imposed Eq.~\eqref{eq:PhiChi}, we see that
\beq
\widetilde{G}(z\rightarrow z_{\rm UV}) = \frac12.
\eeq
We then conclude that the field $A_0$ in this gauge vanishes at the UV boundary, and can be expressed as
\beq
A_0 = G(z)a\bchi\cdot\btau a^{-1},\qquad G(z\rightarrow z_{\rm UV}) = 0.
\eeq
We notice that this result is exactly what one would expect by
allowing for the most general field configuration respecting spherical
symmetry, homogeneity in three-dimensional flat space, and the gauge
choice $A_z=0$.
Taking the functions $H,\wha_0,G,L$ to be independent of $\bchi$
amounts to considering a slow rotation, thus including only linear
terms in $\bchi$ in the Ansatz.

Whereas $\frac12\bchi\cdot\btau$ is the matrix form of the boundary angular
velocity, $\frac12a\bchi\cdot\btau a^{-1}=-\i\dot{a}a^{-1}$ is the
matrix form of the boundary angular isospin velocity (i.e.~describing
rotations in $\SU(2)$ instead of in space).
Thus, although one may think we are spinning the fields in space, this
is really an isospin action on the homogeneous fields.

The final form of our time-dependent homogeneous Ansatz is then
summarized in compact notation as:
\begin{align}
	\calA_0 = G a\bchi\cdot\btau a^{-1} + \tfrac12\wha_0,\quad
	\calA_i = -\tfrac12\left(H a \tau^i a^{-1} + L\chi^i\right),\quad
	\calA_z = 0,
	\label{eq:Ansatzchi}
\end{align} 
with the mandatory boundary condition $G(z\rightarrow z_{\rm UV})=0$.

This Ansatz leads to the action
\begin{align}
  S_{\rm YM} = &-\kappa\int\d^4x\int\d z\bigg[
    -8hH^2\left(G+\frac12\right)^2\bchi\cdot\bchi
    +3hH^4\\
    &+k\left[(L')^2  - 4(G')^2 + 8(KH)^2\right]\bchi\cdot\bchi
    +3k(H')^2
    -k(\wha_0')^2
    \bigg],\non
  S_{\rm CS} = &-\frac{N_c}{8\pi^2}\int\d^4x\int\d z\;\wha_0H'H^2+\frac{N_c}{4\pi^2}\int\d^4x\int\d z
  \left(LH'-L'GH\right)H\bchi\cdot\bchi,
\end{align}
which gives rise to the equations of motion
\begin{align}
	h H^3 - \frac12\p_z(k H') - \frac{N_c}{32\pi^2\kappa}H^2\wha_0' &= 0,\\
	\p_z(k\wha_0') + \frac{3N_c}{16\pi^2\kappa}H^2H' &= 0,\\
	\p_z(k G') - h H^2(1 + 2G) + \frac{N_c}{32\pi^2\kappa} H^2L' &= 0,\label{eq:eomG}\\
	\p_z(k L') + \frac{N_c}{8\pi^2\kappa} H\left[H G' + (1 + 2G)H'\right] &=0,\label{eq:eomL}
\end{align}
where the first two equations of motion are truncated to order
$|\bchi|^0$, whereas the latter two only appear at quadratic order in $\bchi$.
Including the subleading $|\bchi|^2$ corrections to the solutions of $H$
and $\wha_0$ has a negligible impact, which we checked explicitly. 
Moreover, in the limit of small $\bchi$, quadratic corrections in $\bchi$
to the functions $H(z),\wha_0(z)$ would generate terms in the on-shell action
at order $|\bchi|^4$, not contributing to the symmetry energy, and also being subleading
in the small $\bchi$ expansion.

This set of equations is composed by ODEs in the holographic
coordinate $z$ and can be solved with standard off-the-peg solvers in
packages like \textsc{Mathematica} or \textsc{Matlab}, once we specify
all the boundary conditions:
\begin{equation}
    G'(z_{\rm IR})=-\frac{N_c}{32\pi^2\kappa}H^2(z_{\rm IR})L(z_{\rm IR}),\qquad
    \wha_0'(z_{\rm IR})=L'(z_{\rm IR})=0, \qquad
  H(z_{\rm IR})=\epsilon(4\pi^2\rho)^{\frac13},
  \label{eq:IR_BC}
\end{equation}
and all fields are vanishing at $z=z_{\rm UV}$.
These boundary conditions are obtained by imposing the vanishing of
also the total derivative that comes about when deriving the
Euler-Lagrange field equations; for more details, see
Ref.~\cite{Bartolini:2023eam}.\footnote{
Imposing the coupled Robin-type boundary condition for $G'$ as
opposed to a Neumann boundary condition (a naive but consistent
choice based on the field's parity if $L(z_{\rm IR})=0$) only
leads to a decrease in the symmetry energy of about 10-20\%.}
We recall that that in the chosen coordinates
$z_{\rm IR}=0$ ($z_{\rm IR}=1$),
$z_{\rm UV}=\infty$ ($z_{\rm UV}=0$) and
$\epsilon=+1$ ($\epsilon=-1$) for the WSS (HW) model.

Another possible approach would be to keep the fields in a
static configuration, hence keeping the freedom to set the standard
orientation of Eq.~\eqref{eq:Ansatz}, and introduce an external
isospin chemical potential, which holographically amounts to
introducing a finite UV boundary value for the field $A_0$:
in App.~\ref{A:mu} we show that
this approach is related to ours by a gauge transformation, hence
leading to the same physics. This formalism is the one employed in
\cite{Kovensky:2021ddl,Kovensky:2023mye}: the two calculations, however, differ 
in that in the present work we have turned on the Abelian field $\widehat{A}_i$, 
which turns out to be linear in $\bchi$, and we are effectively truncating the $\bchi$
dependence of the gauge fields at linear order. The inclusion of the Abelian
component is necessary to have a self-consistent Ansatz, as the equations 
of motion cannot be solved by setting $L(z)=0$ (the Chern-Simons term provides
a source for $L(z)$). Moreover, it turns out that the field $L(z)$
dominates the small-$\lambda$ behavior of the symmetry energy:
despite the holographic model 
being developed with the large-$\lambda$ limit in mind, for the
practical application of extracting a value for the symmetry energy, we
need to extrapolate to a finite-$\lambda$, and the most popular fit of
the model employs the value of $\lambda=16.63$, which does not realize
the large-$\lambda$ nor the small-$\lambda$ regimes (see App.~\ref{app:large_lambda_approx}). 
On top of the difference at the level of the Ansatz, another
difference with respect to Refs.~\cite{Kovensky:2021ddl,Kovensky:2023mye} lies in the implicit definition of the isospin number of nucleon states in the large-$N_c$ limit.
We choose as proton (neutron) state the lowest-lying isospin state,
which can be thought of as being composed of
$\frac{1}{2}\left(N_c+1\right)$ up (down) and
$\frac{1}{2}\left(N_c-1\right)$ down (up) quarks.
With this definition, the angular velocity $\bchi$ of a nucleon state
is of order $N_c^{-1}$, and so are the isospin chemical potential and
the symmetry energy\footnote{Note that the symmetry energy is a
$1/N_c^2$ correction to the leading $\mathcal{O}(N_c)$ baryon energy,
while corrections from the axial anomaly would be further suppressed
as $1/N_c$ and can provide corrections to the symmetry energy only at
order $\mathcal{O}(N_c^{-2})$, see App.~\ref{app:chiralanomaly}.}.
A different choice that still reduces to the familiar $N_c=3$ case is
that in which the proton (neutron) is composed of $N_c-1$ up (down)
and one down (up) quarks: in this scenario the isospin number is of
order $N_c$, and so is the symmetry energy. We find appropriate the
former definition for nucleon states, in that it keeps the nucleons as
the ground state baryons, and preserves the familiar electric charge
following the Gell-Mann-Nishijima formula
\beq
Q=I_3 + \frac{N_B}{2},
\eeq
with $Q,I_3,N_B$ being the electric charge, the third component of the
isospin, and the baryon number, respectively.

The truncation of the $\bchi$ dependence (and so of the dependence on
$\mu_I$) is an approximation that does not affect the computation, 
since the symmetry energy is by definition obtained by evaluating the
first nonvanishing term in the expansion of the energy per nucleon
around an isospin symmetric configuration.

Despite this technique being equivalent to the usual introduction of 
a boundary chemical potential $\mu_I$, it has a series of advantages, particularly
manifest in the small $\mu_I$ limit. In this limit, the complicated picture
of the isospin asymmetric homogeneous matter becomes similar to the 
well understood one of a slowly rotating bulk instanton, and the problem of 
finding the symmetry energy becomes the computation of a moment of
inertia.

It is also built-in in this formalism what the smallest value of the
isospin is. Identifying this smallest unit of isospin with that of a
single quark being flipped from down to up fixes the isospin quantum
number without $N_c$-scaling ambiguities.

On top of this simplification, our alternative technique also helps in identifying 
the solution to the problem of the ambiguity of the Chern-Simons term when
dealing with homogeneous nuclear matter. As pointed out in Ref.~\cite{Bartolini:2023eam}, 
boundary terms arising from the Chern-Simons term can contribute with an 
IR effective action because of the discontinuity of $H(z)$. In this case, different
choices for the Chern-Simons action that differ by a boundary term are not physically
equivalent (as opposed to the case of a smooth instanton), as they enforce
different IR boundary conditions on the flavor fields, with the
consistent case giving rise to the boundary conditions \eqref{eq:IR_BC}.

A way of solving the ambiguity is to require that the holographic currents on the 
boundary match with their sources in the bulk: in Ref.~\cite{Bartolini:2023eam} this 
was done for the baryonic density $\rho$ (requiring that the topological charge matches 
with the baryonic current in the tail of $\wha_0$) and for the isospin density $\rho_I$
(requiring that the angular velocity description matches with the
isospin chemical potential one). 

This kind of problem arises in every holographic model containing a Chern-Simons term in the
action, hence this result obtained with the aid of the new quantization technique is very 
generalizable and will prove useful in identifying the correct Chern-Simons action for
future works exploring isospin asymmetry in holographic models. The correct choice of the 
action is crucial to obtain the correct thermodynamic quantities, so an improvement in
this regard directly translates to a more precise equation of state,
and more reliable predictions for properties of neutron stars.

\section{Symmetry energy}

The terms quadratic in $\bchi$ exactly produce the SE upon Hamiltonian
quantization:
\begin{align}
  H &= \frac12V\Lambda\bchi\cdot\bchi + VU\non
  &= 2V\Lambda \dot{a}_m^2 + VU\non
  &= \frac{\pi_m^2}{8V\Lambda} + VU \non
  &= \frac{I(I+1)}{2V\Lambda} + VU,
\end{align}
where canonical quantization of $a_m$, $m=0,1,2,3$, a coordinate on the
3-sphere ($a_m^2=1$), leads to the momentum conjugate
\beq
\pi_m=\frac{\p H}{\p\dot{a}_m}=4V\Lambda\dot{a}_m,
\eeq
and hence to $\pi_m^2=\ell(\ell+2)$ being the spherical harmonics and
$\ell=2I$, with $I$ the isospin quantum number
\cite{Adkins:1983ya}\footnote{Due 
to the simplicity of the homogeneous Ansatz, the isospin quantum
number is identical to the spin quantum number in magnitude; this is
an artifact of the Ansatz, but it does not increase the kinetic
energy. In particular, for reading off the coefficient of the symmetry energy
at $\beta=0$, this artifact of the approximation of using the
homogeneous Ansatz is irrelevant. }.
The identification of $V\bchi^2$ and $I(I+1)/V$ coming from Hamiltonian
quantization is also justified by the holographic dictionary, since it
can be obtained by computing the third component of the isovector
charge density, see App.~\ref{app:isovector} for a detailed
computation.
The functionals $\Lambda$ and $U$ are defined as
\begin{align}
  \Lambda &= 2\kappa
  \int\d z\left[2h H^2(2G+1)^2 + k((L')^2 + 4(G')^2)\right],\non
  U &= \kappa
  \int\d z\left[3h H^4 + 3k(H')^2 + k(\wha_0')^2\right],
\end{align}
where $V$ denotes the spatial 3-volume.
Using now the relation between isospin and the number of protons and
neutrons:
\beq
2I=Z-N=-\beta A,
\eeq
with $Z$ the proton number and $N$ the
neutron number, as well as the atomic number
\beq
A=Z+N=V\rho,
\eeq
being the product of the 3-volume and the baryonic density.
$\beta$ is defined as the normalized difference between the number of
neutrons and protons, $\beta=(N-Z)/A$, hence we have
\begin{align}
\frac{H}{A} &= \frac{U}{\rho} + S(\rho)\beta^2
+ \mathcal{O}(V^{-1}),\\
S(\rho) &= \frac{\rho}{8\Lambda},
\end{align}
where $S(\rho)$ is the symmetry energy as a function of the density.

Using the standard phenomenological fit for the WSS model of
Ref.~\cite{Sakai:2005yt}, we set $\lambda=16.63$ and find the first SE
expansion parameters as
\begin{align}
S_0 &= 74.9\left(\frac{M_{\rm KK}}{949\MeV}\right)\MeV,\non
L &= 113.3\left(\frac{M_{\rm KK}}{949\MeV}\right)\MeV,\\
K_{\rm sym} &= -35.9\left(\frac{M_{\rm KK}}{949\MeV}\right)\MeV,\nonumber
\end{align}
which are somewhat larger than values typically obtained from
phenomenological models \cite{Baldo:2016jhp}, but much smaller than
obtained in the WSS previously \cite{Kovensky:2021ddl}.
For the HW model, we fix $M_5=\frac{N_c}{12\pi^2}$ using the leading
OPE coefficient of the vector current correlator \cite{Erlich:2005qh},
for which the first few SE expansion parameters are
\begin{align}
S_0 &= 70.4\left(\frac{\sfL^{-1}}{150\MeV}\right)\MeV,\non
L &= 132.5\left(\frac{\sfL^{-1}}{150\MeV}\right)\MeV,\\
K_{\rm sym} &= -218.8\left(\frac{\sfL^{-1}}{150\MeV}\right)\MeV,\nonumber
\end{align}
where $\sfL^{-1}$ is the mass scale of the HW model, which we set as
$\sfL^{-1}=150\MeV$ following Ref.~\cite{Bartolini:2022rkl}, which provides
phenomenologically good results for neutron stars in terms of
mass-radius data.
The saturation density, $\rho_0$, in HQCD is defined to be at the onset
of baryonic matter, obtained by minimization of the free energy with
the baryonic chemical potential as the boundary condition
$\widehat{a}_0(z_{\rm UV})=\mu_B$.
The value of $\rho_0$ obtained this way in the WSS model with
$\lambda=16.63$ is 
\beq
\rho_0 = 0.436\left(\frac{M_{\rm KK}}{949\MeV}\right)^3\fm^{-3},
\eeq
which is about $2.9$ times too large with respect to the
phenomenological value -- an overestimate by the same order of
magnitude as the other baryonic quantities.
The HW model, however, yields a more realistic saturation density
\beq
\rho_0 = 0.183\left(\frac{\sfL^{-1}}{150\MeV}\right)^3\fm^{-3},
\eeq
which is only about $22\%$ too large.

\begin{figure}[!htp]
  \includegraphics[width=0.49\linewidth]{{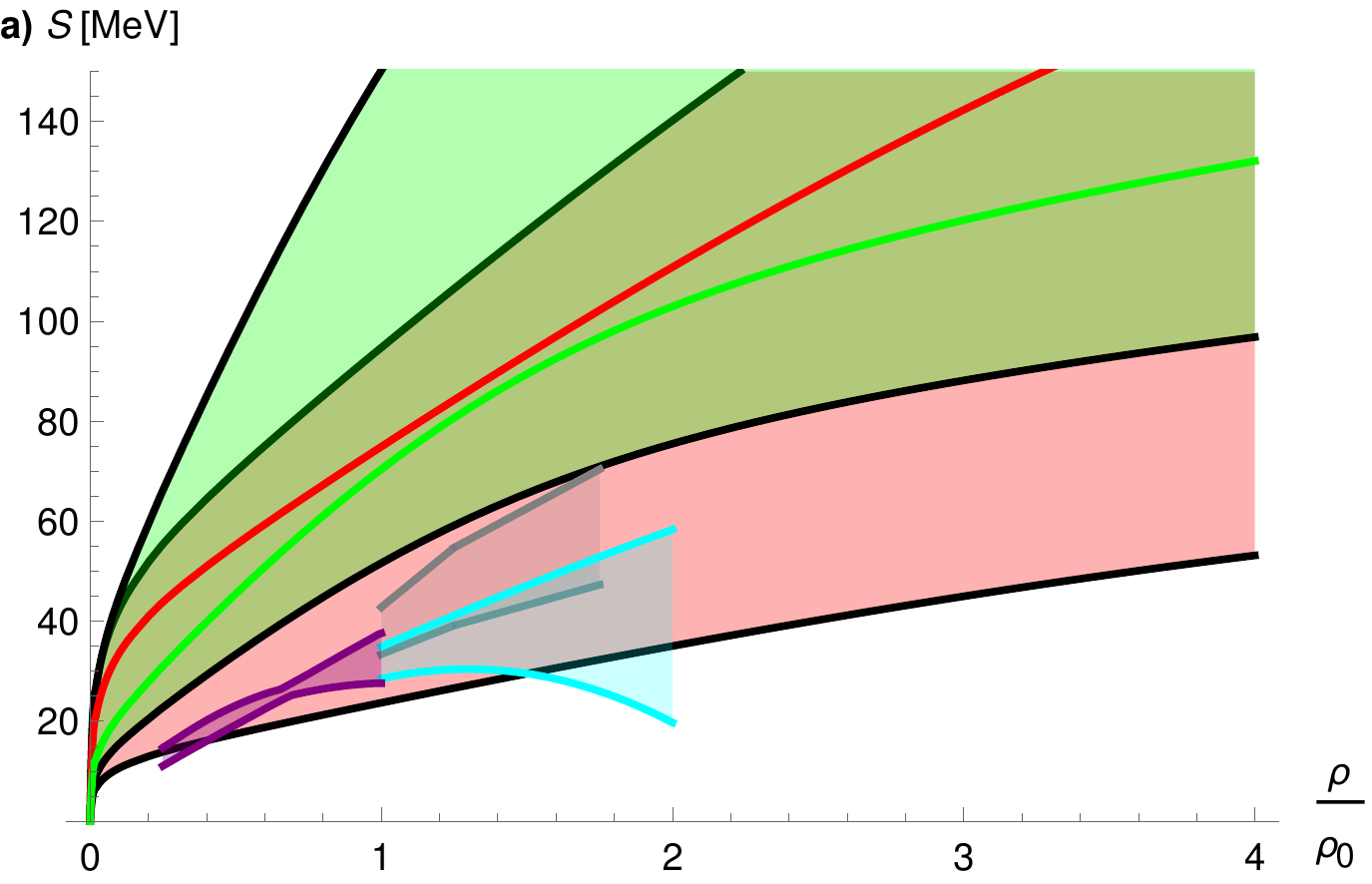}}
   \includegraphics[width=0.49\linewidth]{{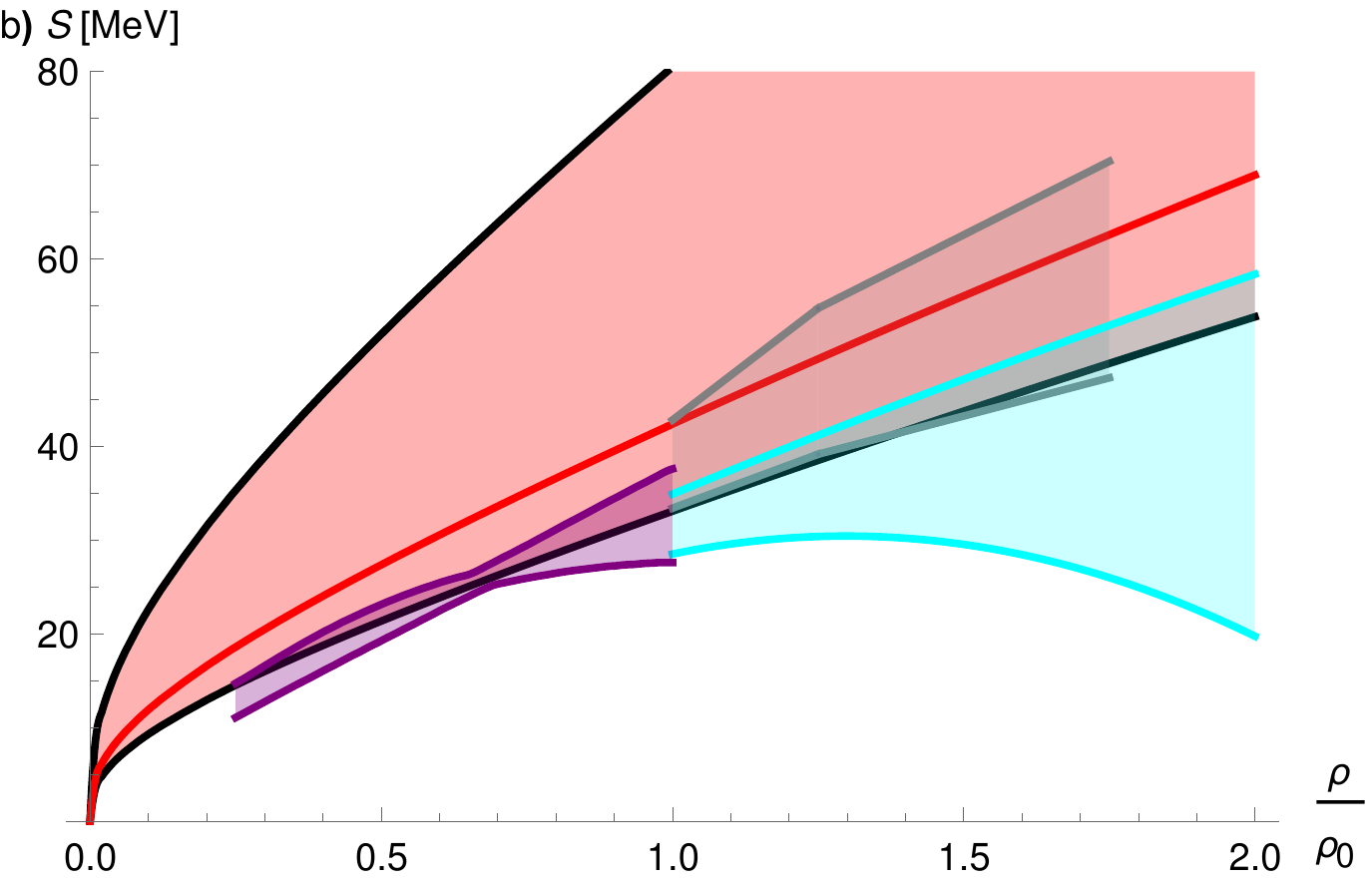}}
  \caption{The symmetry energy (SE) calculated in the WSS model with the
    phenomenological value of the 't Hooft coupling and in the HW
    model, both using quantization of isospin as functions of the
    density.
    a) The red area corresponds to the WSS model with
    $M_{\rm KK}$ ranging from $300\MeV$ to $1200\MeV$ and the red
    line in the middle is at $949\MeV$.
    The green area corresponds to the HW model with $\sfL^{-1}$
    ranging from $110\MeV$ to $320\MeV$ and the green line in the
    middle is at $150\MeV$. 
    The constraints from the PREX-II experiment using the neutron skin
    thickness of $^{208}$Pb \cite{Reed:2021nqk} are shown with a gray
    shaded area, while the extensive 2021 survey of constraints on the
    symmetry energy of Li et.al.~\cite{Li:2021thg} using neutron stars,
    are shown with a cyan shaded area.
    Constraints from isobaric analog states below saturation density
    are shown with a purple shaded area \cite{Danielewicz:2013upa}.
    b) The SE calculated in the WSS model with the 't Hooft coupling
    $\lambda=60$.
    The red shaded area corresponds to $M_{\rm KK}\in[390,949]\MeV$
    and the red solid curve is the rescaled phenomenological mass
    scale, that keeps the pion decay constant at $93\MeV$,
    corresponding to $M_{\rm KK}=500\MeV$.
  }
  \label{fig:esymrho}
\end{figure}
We explore a larger range of densities for both the WSS and the HW
model in Fig.~\ref{fig:esymrho}a). For the WSS model, we have used the
phenomenological value of the 't Hooft coupling ($\lambda=16.63$) and
shown the range of $M_{\rm KK}\in[300,1200]\MeV$ with a red shaded
area, which includes $M_{\rm KK}=949\MeV$ \cite{Sakai:2005yt} (red
solid line), whereas
for the HW model the range of $\sfL^{-1}\in[110,320]\MeV$ is shown
with a green shaded area, which includes $\sfL^{-1}=150\MeV$ (green
solid line) that is chosen from neutron star phenomenology
\cite{Bartolini:2022rkl} and the highest mass scale is from meson
physics \cite{Ilderton:2005yu,DaRold:2005vr}.
Up-to-date constraints from astrophysics and heavy-ion collision
data are shown with gray and cyan shaded areas near and above
saturation density and constraints using nuclear excitation energies
from isobaric analog states (IAS) are shown with a purple shaded area
below saturation density.
As can be seen from the figure, the phenomenologically fitted value of
the mass scale $M_{\rm KK}$ at $949\MeV$ \cite{Hata:2007mb}
overestimates the SE with about a factor of $2.4$; however, the fit is
made using mesonic observables and is known to overestimate baryonic
observables; for instance, the baryon mass is typically overestimated
by a factor of 1.7-1.8
\cite{Hata:2007mb,Baldino:2017mqq,Baldino:2021uie} using the mesonic
fit.

Although both models come in the ball park of the experimental
constraints above saturation density and nuclear physics predictions
below saturation density if we allow ourselves to adjust the
energy scale ($M_{\rm KK}$ or $\sfL^{-1}$), the shape of the SE
does not quite satisfy all the constraints.
In the WSS model, however, we can dial the 't Hooft coupling to see
whether we can fit in the allowed regions and indeed it is possible by
raising both the 't Hooft coupling from the phenomenological value to
$\lambda=60$, as well as lowering the KK scale from $949\MeV$ to
$390\MeV$, see Fig.~\ref{fig:esymrho}b); this corresponds to the
lower black curve of the red shaded area.
With the larger 't Hooft coupling, the SE of the WSS has a compatible
shape to pass all the constraints, but the $\rho$ meson is too light
and the baryon mass is too heavy -- often a problem in HQCD;
the baryon mass is reduced from about $1600\MeV$ to $1191\MeV$.
If we keep the pion decay constant at its phenomenological value, the
KK scale, however, is lowered too and is shown in Fig.~\ref{fig:esymrho}b)
with a solid red curve -- not too far from a viable solution.
Recomputing the symmetry energy expansion parameters, we obtain
\begin{align}
S_0 &= 33.1\left(\frac{M_{\rm KK}}{390\MeV}\right)\MeV,\non
L &= 66.4\left(\frac{M_{\rm KK}}{390\MeV}\right)\MeV,\\
K_{\rm sym} &= -34.3\left(\frac{M_{\rm KK}}{390\MeV}\right)\MeV,\nonumber
\end{align}
which are compatible with phenomenological constraints.
The saturation density for this fit now also has improved as
\beq
\rho_0 = 0.166\left(\frac{M_{\rm KK}}{390\MeV}\right)^3\fm^{-3},
\eeq
which is only about $10\%$ from the phenomenological value.
Lowering the KK scale from 390 to $380\MeV$ will improve both $S_0$
and $\rho_0$ ($S_0=32.2\MeV$ and $\rho_0=0.153\fm^{-3}$), but will
create a bit more tension with the constraint coming from the neutron
skin thickness of ${}^{208}$Pb, shown with a gray-shaded area in
Fig.~\ref{fig:esymrho}b). A similar choice of fit is employed in 
Ref.~\cite{Bartolini:2023wis}, where we choose $M_{\rm KK}$ and $\lambda$
in order to reproduce the correct saturation density and SE. The resulting
equation of state (hybridized with phenomenological low-density EOSs)
then is found to reproduce viable properties of neutron stars.

It is not surprising that a somewhat larger value of $\lambda$ is needed
in order to more reliably reproduce the physics of baryonic matter. We can
understand it by considering the single baryon in the WSS model: with the BPST 
instanton approximation it is possible to compute both the classical mass and 
its quantum corrections \cite{Hata:2007mb}. In particular, since we are interested
in the symmetry energy, we want to consider the (iso)spin quantum correction, given by
integer values of $l$ in the formula \cite{Hata:2007mb}
\beq
M= 8\pi^2 \kappa +\sqrt{\frac{(l+1)^2}{6}+\frac{2}{15}N_c^2} + \frac{2(n_\rho + n_Z +1)}{\sqrt{6}},
\eeq
where $n_\rho,n_Z$ are quantum numbers for the size and bulk position excitations.

It is well known that the usual mesonic fit with $M_{\rm KK}=949\MeV$, $\lambda=16.63$ largely 
overestimates the nucleon masses, but a major contribution in this result comes from the fact that
the quantum corrections are close in magnitude to the classical mass.
We can see that the classical mass is of order $\calO(\lambda)$, while the quantum corrections are 
of order $\calO(1)$.
By increasing $\lambda$ it is then possible to reduce the relative magnitude of the quantum
corrections as compared to the classical result, and by reducing
$M_{\rm KK}$ it is possible to obtain an overall more realistic
nucleon mass. 

The same kind of mechanism is inherited by the homogeneous system,
where increasing $\lambda$, the symmetry energy contribution becomes
smaller, moving towards the real world value, provided that we also
adjust $M_{\rm KK}$ accordingly. While it is fairly easy to expect
that some values of $\lambda$, $M_{\rm KK}$ exist that both fit the
phenomenology of saturation density and symmetry energy, it is less
trivial that once these values are employed, other baryonic
observables are then improved, as it happens in our case with the BPST
instanton mass and with the expansion parameters $L$, $K_{\rm sym}$.

What this analysis suggests is that when describing baryons within the
WSS model, the errors introduced by the many approximations employed
to make the system approachable seems to be partially mitigated by an
alternative choice of the values of the parameters $\lambda$,
$M_{\rm KK}$. 

\begin{figure}[!tp]
  \centering
  \includegraphics[width=0.5\linewidth]{{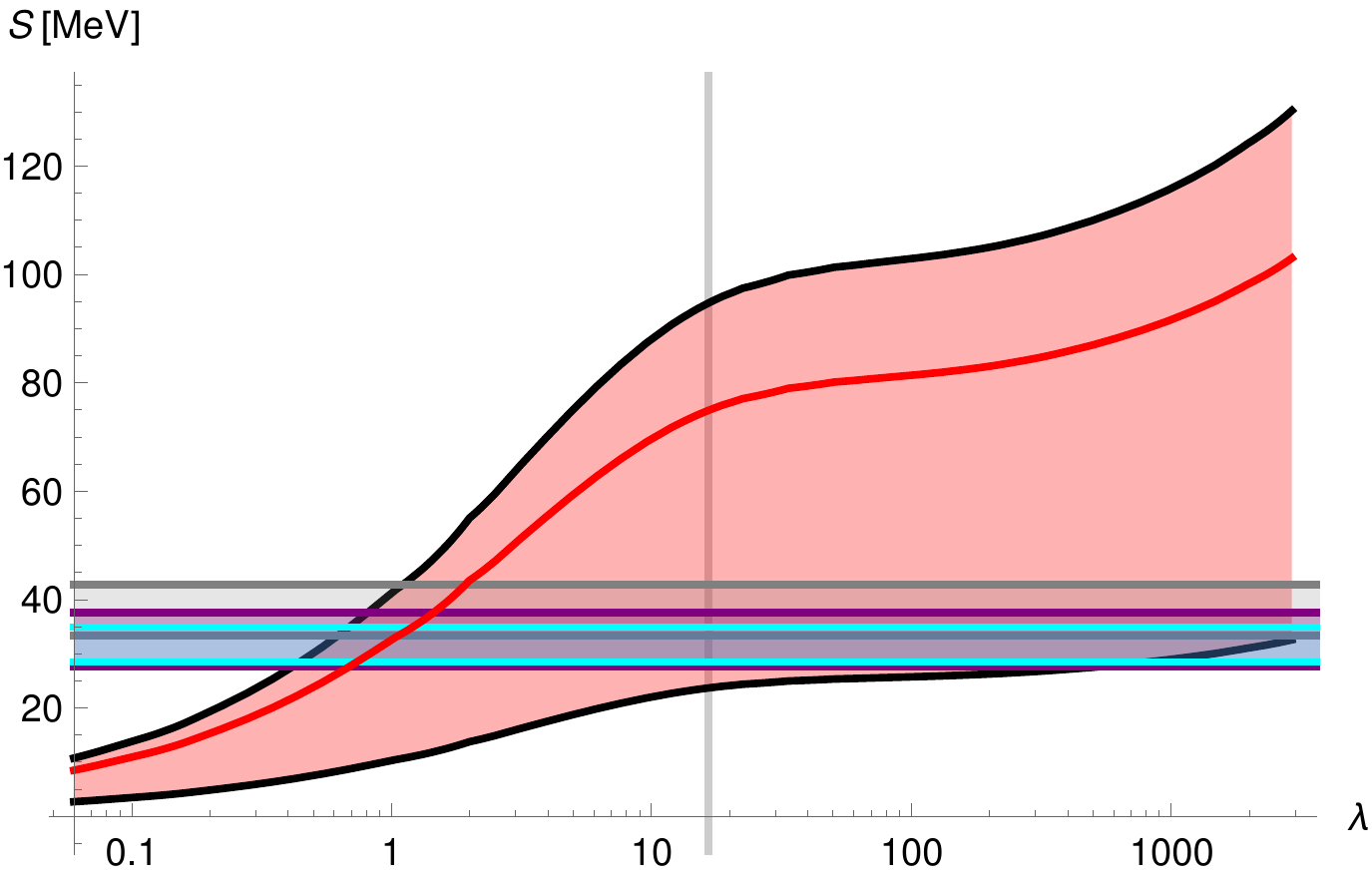}}
  \caption{The symmetry energy (SE) calculated in the WSS model using
    quantization of isospin as a function of the 't Hooft coupling
    $\lambda$ at saturation density.
    The red area spans the mass scale $M_{\rm KK}$ from $300\MeV$ to
    $1200\MeV$ and the red line is at $949\MeV$. The gray, cyan and purple
    shaded areas are the same as in Fig.~\ref{fig:esymrho}.
    The vertical gray line marks the phenomenological 't Hooft
    coupling $\lambda=16.63$ \cite{Sakai:2005yt}.}
  \label{fig:esymlambda}
\end{figure}
The dependence on the 't Hooft coupling for the WSS model is shown in
Fig.~\ref{fig:esymlambda} for the KK scale in the interval
300-1200$\MeV$ at saturation density.

\section{Proton fraction}

We will now consider the proton fraction at $\beta$-equilibrium with
charged leptons, imposing charge neutrality.
Using the Gell-Mann-Nishijima formula, we can relate the baryon
density, $\rho$, and isospin density, $\rho_I$, with the
proton/neutron densities:
\beq
\rho_{P,N}=\tfrac12\rho\pm \rho_I,
\eeq
where the upper sign is for protons and the lower for neutrons.
Charge neutrality is imposed by
\beq
\tfrac12\rho+\rho_I = {\textstyle\sum}_\ell\rho_\ell,
\label{eq:charge_neutral}
\eeq
with $\ell=e,\mu$ being a sum over the charged leptons and the
$\beta$-equilibrium (from the decay $N\to P+\ell+\bar{\nu}_\ell$)
amounts to
\beq
\mu_\ell = \mu_N - \mu_P = -\mu_I, \qquad
\ell=e,\mu,
\label{eq:beta_equilibrium}
\eeq
where $\mu_X$ is the chemical potential of the particle species $X$.
The lepton density is calculated assuming it to be a (massive) Fermi
gas as \cite{Kovensky:2021ddl}
\beq
\rho_\ell = \Theta_H(\mu_\ell-m_\ell)\frac{(\mu_\ell^2-m_\ell^2)^{\frac32}}{3\pi^2},
\label{eq:Fermi_rho}
\eeq
with $\Theta_H$ being the Heaviside step function, $\mu_\ell$ the
chemical potential and $m_\ell$ being the mass of the lepton $\ell$.
Using the definition of the isospin chemical potential as being the
conjugate variable of the isospin density, we get
\beq
\mu_I = \frac{1}{V}\frac{\p H}{\p\rho_I}
= \frac{\rho_I}{\Lambda(\rho)}.
\eeq
Inserting Eq.~\eqref{eq:Fermi_rho} into the charge neutrality
condition \eqref{eq:charge_neutral} and using the $\beta$-equilibrium
condition \eqref{eq:beta_equilibrium}, we obtain an implicit solution
for the isospin density, $\rho_I$, as a function of the density $\rho$:
\begin{equation}
\frac{\rho_I^3}{3\pi^2\Lambda^3}
\left[\Theta_H(-\rho_I)
  +(1-R^{-2}m_\mu^2)^{\frac32}\Theta_H(-R-m_\mu)
  \right]
\mathop+\rho_I + \frac12\rho = 0, \quad
R = \frac{\rho_I}{\Lambda},
\end{equation}
where we have set the electron mass to zero and
the dimensionless muon mass parameter is the ratio of the
physical mass ($105.7\MeV$) to the mass scale $M_{\rm KK}$ and
$\sfL^{-1}$, for the WSS and the HW models, respectively.

\begin{figure}[!tp]
  \centering
  \includegraphics[width=0.5\linewidth]{{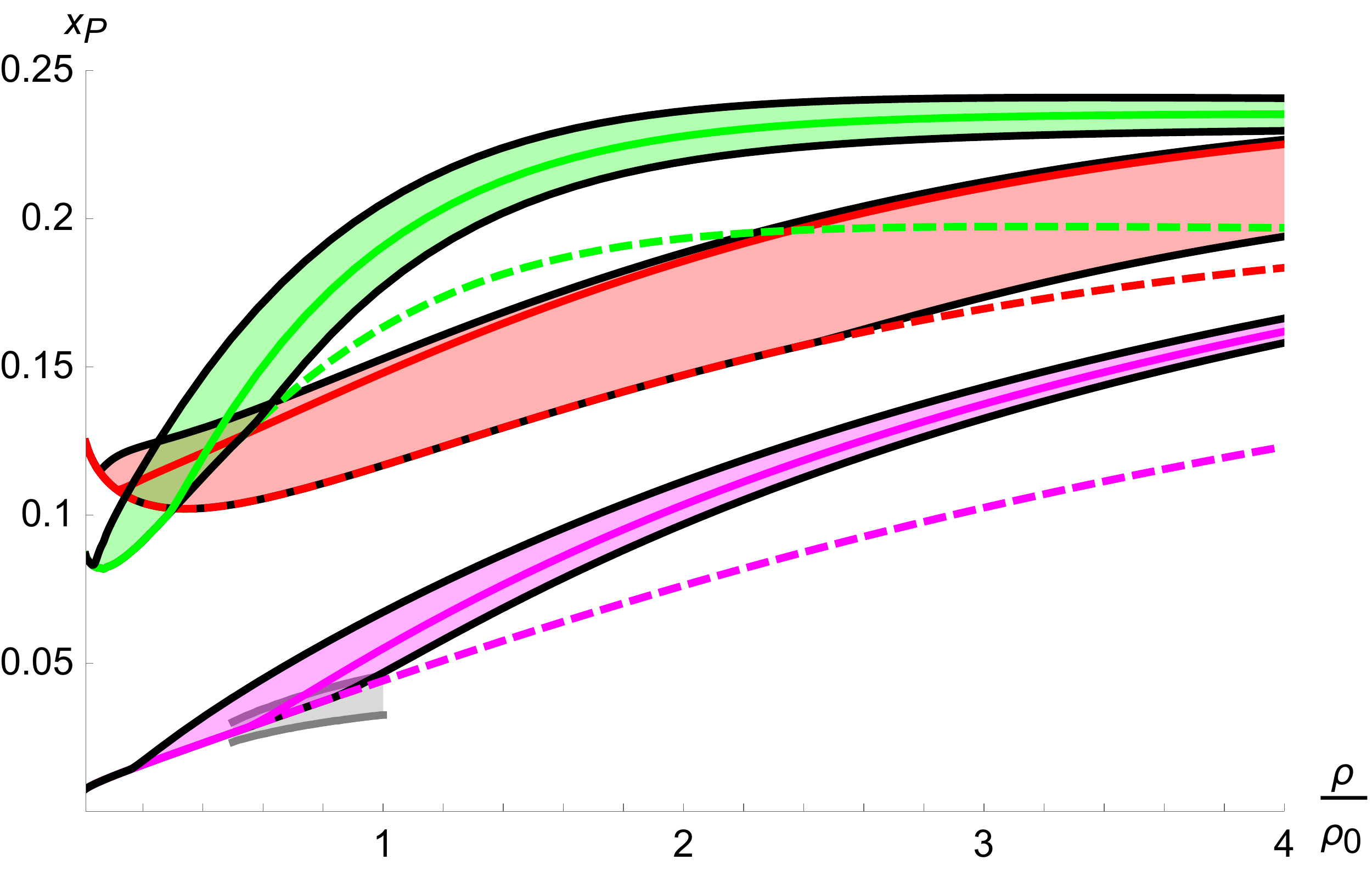}}
  \caption{The proton fraction calculated in the WSS and HW model as
    functions of the density.
    The red shaded area corresponds to the WSS model with the
    phenomenological 't Hooft coupling $\lambda=16.63$ and
    $M_{\rm KK}\in[300,1200]\MeV$, the red line at
    $M_{\rm KK}=949\MeV$ and the red dashed line at $M_{\rm KK}\to0$.
    The green shaded area corresponds to the HW model with
    $\sfL^{-1}\in[110,320]\MeV$, the green line at $\sfL^{-1}=150\MeV$
    and the green dashed line is the limit $\sfL^{-1}\to0$.
    The magenta shaded area corresponds to the WSS model with the
    calibration of Fig.~\ref{fig:esymrho}b), i.e.~$\lambda=60$ and
    $M_{\rm KK}\in[390,949]\MeV$, the solid magenta curve corresponds
    to the phenomenological pion decay constant, and the dashed
    magenta curve corresponds to $M_{\rm KK}\to0$, thus eliminating
    the muons.
    The gray shaded area is the result from chiral EFT \cite{Lim:2017aac}.
    }
  \label{fig:protonfrac}
\end{figure}
In Fig.~\ref{fig:protonfrac} we show the numerical results for both
the WSS and the HW model for the proton fraction at various densities
around saturation density.
We find that the
WSS model phenomenologically fitted to mesons gives more realistic
proton fractions (red shaded area) than the HW model (green shaded
area) and yields even better proton fractions below saturation
density if we use the calibration from Fig.~\ref{fig:esymrho}b),
i.e.~$\lambda=60$ and $M_{\rm KK}=390\MeV$.
Since we take the electrons to be massless, the mass scale of the
model only enters in the muon mass parameter.
The dashed curves correspond to the muon being infinitely heavy (or
the mass scale of the model being sent to zero).

\section{Discussion and outlook}

In this paper, we have computed the symmetry energy in two
holographic QCD models using the method of quantizing the isospin
symmetry, namely in the top-down WSS model and the
bottom-up HW model.
We find fairly good agreement between our model results for the SE and
proton fraction in both the HW model, using the fit from neutron stars
and in the WSS model with a new calibration (i.e.~$\lambda\sim60$ and
$M_{\rm KK}\sim390\MeV$).

We have also shown that the method known from Skyrmions of quantizing
the isospin zeromode is equivalent to introducing a chemical potential
on the holographic boundary for the gauge fields, see App.~\ref{A:mu}.
There is mathematically no difference between the two methods.

It would be interesting in future work to take into account the
strange quark (3 instead of 2 flavors) or alternatively the kaons, to
see at what densities it might have an impact on the SE.
Furthermore, there are certain transitions that happen at larger
densities, for example the Skyrmion-half-Skyrmion transition
\cite{Park:2009mg}, which has an analog in holographic instantons
\cite{Rho:2009ym}.
Although it is not directly observable in our homogeneous Ansatz, it
may have some effect on the SE and proton fractions at large
densities.
Another approximation we employed in our calculations is that of keeping
fixed the position of the discontinuity that sources the baryonic charge.
It is expected that the location of the baryon is dynamically determined, moving 
towards the boundary as the density is increased: this behavior is the
(homogeneous version of) the so-called ``popcorn transition'' that is known
to occur in HQCD \cite{Kaplunovsky:2012gb}. 
However, as shown in Ref.~\cite{Bartolini:2022rkl}, the homogeneous Ansatz already
reproduces features of the popcorn transition even when keeping the position
of the discontinuity fixed in the bulk: it does so by having the baryonic holographic density 
form a peak at a location in the bulk that depends on the boundary density $\rho$.
Because of this, we expect corrections coming from the dynamical determination of the 
discontinuity location to be particularly small.

\subsection*{Acknowledgments}
We thank Nicolas Kovensky, Anton Rebhan and Andreas Schmitt for
discussions and comments on the draft.
The work of L.~B.~is supported by the National Natural Science
Foundation of China (Grant No.~12150410316). 
S.~B.~G.~thanks the Outstanding Talent Program of Henan University and
the Ministry of Education of Henan Province for partial support.
The work of S.~B.~G.~is supported by the National Natural Science
Foundation of China (Grants No.~11675223 and No.~12071111) and by the
Ministry of Science and Technology of China (Grant No.~G2022026021L).

\appendix
\section{Equivalence between rotation in
  \texorpdfstring{$\SU(2)$}{SU(2)} and external isospin chemical
  potential}\label{A:mu}
We start with our field configuration given by
Eq.~\eqref{eq:Ansatzchi}, of which we rewrite the non-Abelian
components: 
\begin{align}
	A_0 &= G a \bchi\cdot\btau a^{-1} \\
	A_i &= -\frac{H}{2}a\tau^i a^{-1} ,\\
	A_z &= 0.
\end{align} 
We now perform a gauge transformation, with the aim of obtaining a
static configuration in the limit of constant $\bchi$: we choose a
gauge function $b(t)$ that only depends on time, so that the fields
transform as 
\begin{align}
	A_0 \rightarrow &\;\widetilde{A}_0= G b a \bchi\cdot \btau a^{-1}b^{-1} -\i b\p_0b^{-1}  ,\\
	A_i \rightarrow &\;\widetilde{A}_i= -\frac{H}{2}ba\tau^i a^{-1}  b^{-1},\\
	A_z \rightarrow &\;\widetilde{A}_z= 0.
\end{align} 
Now we choose $b=a^{-1}$, hence rotating the fields $A_i$ back to the
standard orientation, while modifying the field $A_0$ with an
additional term:
\begin{align}
  \widetilde{A}_0 &= G  \bchi\cdot\btau -\i a^{-1}\dot{a}\label{eq:muA0} \\
  \widetilde{A}_i &= -\frac{H}{2}\tau^i  ,\\
  \widetilde{A}_z &= 0.
\end{align} 
We recognize the quantity of Eq.~\eqref{eq:PhiChi} in the last term of
Eq.~\eqref{eq:muA0}: 
\beq
-\i a^{-1}\dot{a}= \frac12 \bchi\cdot\btau,
\eeq
so that we are left with
\beq
\widetilde{A}_0=\left(G+\frac12\right)\bchi\cdot\btau.
\eeq
We know that by construction the function $G(z)$ vanishes at the
boundary at $z_{\rm UV}$, so we conclude that this configuration
behaves as: 
\beq
\widetilde{A}_0 (z\rightarrow z_{\rm UV})= \frac12 \bchi\cdot \btau.
\eeq
The boundary value of the field $A_0$ is dual to an isospin chemical
potential in the holographic dictionary. Since the orientation of
the soliton is a zeromode, we can set $\bchi$ to point in a chosen
direction for simplicity without loss of generality: we choose it to
have only a nonvanishing third component as
$\bchi=\left(0,0,\mu_I\right)$, following the same convention of
choosing the third component of isospin as the operator to diagonalize
simultaneously with the isospin squared (and the isospin chemical
potential to appear holographically as the boundary value of
$A_0^{a=3}$).
Shifting $G(z)$ as
\beq
\widetilde{G}(z) = \left(G(z)+\frac12\right),
\eeq
we obtain the familiar expressions for the gauge field and its boundary condition
\beq
\widetilde{A}_0 = \widetilde{G}\tau^3\mu_I,\qquad
\widetilde{A}_0(z\rightarrow z_{\rm UV}) = \frac12\mu_I\tau^3.
\label{eq:SchmittGauge}
\eeq

We then conclude that a static system in the presence of an external
isospin chemical potential $\mu_I$ is equivalently described as it
rotating in isospin space with angular velocity
$\chi^i = \mu_I \delta^{i3}$, as observed in Ref.~\cite{Adam:2023cee}
in the non-holographic context of the Skyrme model. 

We want to emphasize that, despite our solution to the system of
coupled equations of motion is performed in the limit of small angular
velocity (small $\mu_I$), the equivalence between the two methods just
shown holds true in general, since we made no assumptions on the
$\bchi$ dependence of the functions $H,\wha_0,G,L$. The assumption of
small $\bchi$ will not affect in any way the calculation of the
symmetry energy, as it is the coefficient of a term of an expansion
around symmetric matter, hence all the functions would have to be
evaluated at vanishing isospin density (and $\mu_I$) anyway.
For calculations at higher isospin chemical potential, the
equivalence still holds, but since the now large angular velocity $\bchi$
backreacts every field, this framework loses its main advantage of factorizing 
away the dependence on $\bchi$. Moreover, the isospin symmetric Ansatz is not 
consistent anymore, and the function $H(z)$ has to be substituted with a set of
functions $H_i(z)$ \cite{Kovensky:2023mye}:
\beq
A_i = -\frac{H}{2}\tau^i \qquad\Rightarrow\qquad
A_i = -\frac{H_i}{2}\tau^i,
\eeq
where $i$ is not summed over.
Then the problem to be solved is that of five ($H_1(z)=H_2(z)$ because
of residual symmetry) coupled ODEs, with $\bchi$ dependence in each of
them, effectively the same as we would have if we worked with a
boundary chemical potential $\mu_I$.

Despite this not being a necessity for the calculation of the symmetry energy (since by 
definition it is calculated at vanishing $\mu_I$),
when moving to finite isospin density (for example when computing the full phase diagram or 
properties of neutron stars) the more rigorous approach would be to
include the effects above (and possibly non-diagonal terms for $A_i^a$ when
including also quark masses, see Ref.~\cite{Kovensky:2023mye} for the validity limits
of the diagonal approximation), which is equally difficult with both gauge
choices.

\section{Isospin density from holographic current}
\label{app:isovector}

In this section we want to prove that the isospin number density that
we defined from the quantized angular momentum coincides with the
canonical one obtained through the holographic dictionary via the
computation of the isovectorial current.
For simplicity, we will show the proof in the WSS model, so that
$z_{\rm IR}=0$, $z_{\rm UV}=+\infty$, but it holds true in the HW model
too, after substitution of the appropriate quantities. 
As shown in Ref.~\cite{Hashimoto:2008zw}, the vectorial current is
obtained as 
\beq
\mathcal{J}_{V\mu}=-\kappa \left[k(z)\mathcal{F}_{\mu z}\right]^{+\infty}_{-\infty}=-2\kappa \left[k(z)\mathcal{F}_{\mu z}\right]^{+\infty}_{0}.
\eeq
With this quantity we can build the isovectorial charge $Q_V$ of which
we take the third component to coincide with the isospin operator 
\beq
Q_V^{a=3} = I_3= \int d^3x \Tr \left(J_V^0 \tau^3\right) = V \Tr \left(J_V^0 \tau^3\right).
\eeq
We plug in this formula the homogeneous Ansatz \eqref{eq:Ansatzchi}:
\beq
I_3 &&= -2\kappa V \left[G'k(z)\right]^{+\infty}_{0}\chi^i \Tr\left(a \tau^ia^{-1}\tau^3\right)\\ 
&& = -2\kappa V \left[G'k(z)\right]_{z=+\infty}\chi^i \Tr\left(a \tau^ia^{-1}\tau^3\right),
\eeq
where we used the fact that $G'(0)=0$.

The angular velocity $\chi^i$ is related to the angular momentum
operator $J^i$ by the familiar relation involving the moment of
inertia $\Lambda$: 
\beq
\chi^i = \frac{1}{V \Lambda} J^i,
\eeq
and we can exploit the useful relationship between angular momentum
and isospin operators that holds due to the spherical symmetry of the
system: 
\beq
J^i \Tr\left(a \tau^ia^{-1}\tau^a\right) = -2 I_a,
\eeq
so that we are left with:
\beq
I_3 = \frac{4\kappa}{\Lambda}\left[G'k(z)\right]_{z=+\infty} I_3.
\eeq
We see that the validity of this relationship depends on whether the following identification holds true:
\beq
4\kappa \left[G'k(z)\right]_{z=+\infty} = \Lambda.\label{eq:LambdaI3}
\eeq

To prove this relationship, we first notice that the formula for the
current is obtained by differentiating the action with respect to the
UV boundary value of the $A_0$ field, following 
\beq
\delta_{A_0} S = \text{(e.o.m. terms)} +4\kappa V \Tr\left[k(z) A_0'\dA_0\right]_{z=+\infty},
\eeq
where the first term means that we neglect contributions that vanish
by the equations of motion, we evaluate only boundary terms, and we
made use of the boundary condition $A_0'(z=0)=0$.  We decide to employ
the gauge in which the field $A_0$ has a finite boundary value, dual
to the isospin chemical potential, so we take the field to be as in
Eq.~\eqref{eq:SchmittGauge}. With this configuration, the variation of
the action assumes the shape 
\beq
\delta_{A_0} S = \text{(e.o.m. terms)}
+ 2\kappa V \Tr\left[k(z) \widetilde{G}'\tau^3 \tau^3\right]_{z=+\infty} \mu_I \dmu_I,
\eeq
and finally we can compute the derivative
\beq
\frac{\partial S}{\partial \mu_I}
= 4\kappa V \left[k(z) \widetilde{G}'\right]_{z=+\infty} \mu_I.
\label{eq:dSdmu}
\eeq
We can change this result to our usual ``rotating'' gauge by noting
that we have to rename $\mu_I \rightarrow \chi_3$, and that
$\widetilde{G}' = G'$, so that on-shell we obtain
\beq
\frac{\partial S}{\partial \chi_3} = 4\kappa V \left[k(z) G'\right]_{z=+\infty} \chi_3.
\eeq 
We now look at the definition of $\Lambda$: it is nothing but the part
of the energy density that is quadratic in the angular velocity, and
there is no linear term. In this picture, the system is rotating and
there is no chemical potential, so the on-shell action gives the
energy of the system, so that we can write 
\beq
\frac{\partial S}{\partial \chi_3} = V\Lambda \chi_3. \label{eq:dSdchi}
\eeq
Comparing Eqs.~\eqref{eq:dSdmu} and \eqref{eq:dSdchi}, we finally
prove Eq.~\eqref{eq:LambdaI3}.

\section{The subleading order in \texorpdfstring{$N_c$}{Nc} of the chiral anomaly}\label{app:chiralanomaly}

Throughout the main body of this work, we have ignored the presence of
the chiral anomaly of QCD: we expect on general grounds that the
currents dual to the holographic fields $\calA_\alpha$ are conserved,
with the exception of the axial $\U(1)$ current, since the
corresponding symmetry is broken by the chiral anomaly. 
Analogously, we expect the Goldstone boson associated to the axial
symmetry to acquire a finite mass as a consequence of the anomaly.

This holds true in the WSS model, where the mechanism is incorporated
nontrivially from the top-down construction in string theory.
The model includes Ramond-Ramond forms $C_n$ of odd rank $n$: Among these is $C_7$, whose
action, inclusive of a coupling with the flavor branes, reads
\beq
S_{C_7}= -\frac{1}{4\pi}(2\pi\ell_s)^6\int\d C_7\wedge\star\d C_7 + \frac{1}{2\pi}\int C_7\wedge\Tr \calF\wedge \omega_y,
\eeq
where a one-form $\omega_y=\delta(y)\d y$ has been introduced to model
the distribution of the stack of branes in the $y$-direction (by
definition transverse to $z$), extending the otherwise 9-dimensional
integral to the whole 10-dimensional spacetime.

We can write the equation of motion as
\beq\label{eq:C7motion}
\d\star\d C_7=\d\star F_8=\frac{1}{(2\pi \ell_s)^6}\Tr \calF\wedge\delta(y)\d y,
\eeq
and then use Hodge duality
$\star F_8=(2\pi\ell_s)^{-6}\widetilde{F}_2$ to turn
Eq.~\eqref{eq:C7motion} into an anomalous Bianchi identity:
\beq
\d\widetilde{F}_2=\Tr\calF \wedge\delta(y)\d y.
\eeq
This form is gauge invariant if we allow $C_1$ to transform with a
$\U(1)$ transformation of the flavor group:
\beq
\delta_\Lambda\d C_1 = \sqrt{\frac{N_f}{2}}\d\Lambda \wedge
\delta(y)\d y,\qquad \delta_\Lambda \widehat{A}=-\d\Lambda.
\eeq
The implication of this fact is that $\d C_1$ is not a gauge invariant
form, only $\widetilde{F}_2$ is the correct gauge invariant
combination. 
 
This is welcome, since in the model $C_1$ is dual to the $\theta$
angle of QCD as 
\beq
\theta +2\pi k=\int_{S_{\rm UV}^4} C_1.
\eeq
Let us now consider a zeromode for the field $\widehat{A}_z$, dual to
the $\eta'$ meson, such that 
\beq
\int\d z \widehat{A}_z = \frac{2\eta'(x)}{f_\pi},
\eeq
and plug it into the action 
\beq
S_{\widetilde{F}_2}=-\frac{1}{4\pi (2\pi \ell_s)^6}\int\d^{10}x|\widetilde{F}_2|^2.
\eeq
The result is an action that displays a mass term for the $\eta'$: 
\beq
S_{\widetilde{F}_2}=-\frac{\chi_g}{2}\int\d^4x \left(\theta + \frac{\sqrt{2N_f}}{f_\pi}\eta'\right)^2,
\eeq
with the $\eta'$ mass agreeing with the Witten-Veneziano formula 
\beq
m_{\eta'}^2=\frac{2N_f}{f_\pi^2}\chi_g.
\eeq
The topological susceptibility $\chi_g$ and the pion decay constant
are computed in the model (see Ref.~\cite{Sakai:2004cn}):
\beq
\chi_g=\frac{\lambda^3 M_{\rm KK}^4}{4(3\pi)^6}, \qquad
f_\pi = 2\sqrt{\frac{\kappa}{\pi}}.
\eeq
We now recall that the parameter $\kappa$ was defined to be
$\kappa\equiv\frac{\lambda N_c}{216\pi^3}$: this means that the mass
term for the $\eta'$ meson is of order $N_c^{-1}$ (since
$f_\pi^2\propto\mathcal{O}(N_c)$), while the action 
for the flavor fields that we employed in the main body of this work
is of order $N_c$. 
While true that the angular velocity $\chi_i$ itself is of order
$N_c^{-1}$, hence pushing the isospin asymmetric action (proportional
to $\chi^2$) to be of order $N_c^{-1}$, we have to recall that
eventual isospin asymmetric terms will carry factors of $N_c^{-1}$ or
higher also in the $\eta'$ mass term, pushing it to even higher order
in the $N_c^{-1}$ expansion.
Hence it is formally safe to neglect the contribution from the axial
anomaly in the large-$N_c$ scheme of approximation.

\section{The vanishing of the hard-wall tachyon in the baryonic phase}
\label{app:scalardecoupling}

Let us quickly review the setup of the hard-wall model of
Ref.~\cite{Bartolini:2022rkl} that utilizes a scalar field with an IR
potential to dynamically stabilize it.
The metric is given by
\beq
\d s^2 = \frac{\sfL^2}{z^2}\left(\d x^\mu\d x_\mu - \d z^2\right),
\eeq
where $\sfL:=1$ is the curvature scale of AdS$_5$ and set equal to one.
Upon restoring units, energies are multiplied by a physical scale.

For two flavors, we have left and right $\U(2)$ gauge fields,
$\calL_M$, $\calR_M$ and the minimal action \cite{Bartolini:2022rkl}:
\begin{align}
  S &= S_g + S_{\rm CS} + S_\Phi + S_{\rm IR},\\
  S_g &= -\frac{M_5}{2}\int\d^4x\d z\;a(z)
  \Tr\left(\calL_{MN}\calL^{MN}+\calR_{MN}\calR^{MN}\right),\\
  S_{\rm CS} &= \frac{N_c}{16\pi^2}\int\d^4x\d
  z\;\frac14\epsilon^{MNOPQ}
  \bigg[\widehat{L}_M\left(\Tr(L_{NO}L_{PQ}+\frac16\widehat{L}_{NO}\widehat{L}_{PQ}\right)\non
    &\phantom{=\frac{N_c}{16\pi^2}\int\d^4x\d z\;\frac14\epsilon^{MNOPQ}\bigg[\ }
    -\widehat{R}_M\left(\Tr(R_{NO}R_{PQ}+\frac16\widehat{R}_{NO}\widehat{R}_{PQ}\right)\bigg],\\
  S_\Phi &= M_5\int\d^4x\d z\;a^3(z)
  \left[\Tr(D_M\Phi)^\dag(D^M\Phi)-a^2(z)M_\Phi^2\Tr\Phi^\dag\Phi\right],\\
  S_{\rm IR} &= \frac12m_b^2\xi^2 - \lambda_b\xi^4,
\end{align}
where $a(z)=\sfL/z$, the $\U(2)$ gauge field $\calL_M$ is split into
$\SU(2)$ and $\U(1)$ parts as
\beq
\calL_M = L_M^a\frac{\tau^a}{2} + \widehat{L}_M\frac{\mathbf{1}_2}{2},
\eeq
and similarly for $\calR_M$, the field strength for $\calL_M$ is
\beq
\calL_{MN} = \p_M\calL_N-\p_N\calL_M+\i[\calL_M,\calL_N],
\eeq
and similarly for $\calR_{MN}$, the covariant derivative for the
scalar field is defined as
\beq
D_M\Phi = \p_M\Phi + \i\calL_M\Phi - \i\Phi\calR_M,
\eeq
the boundary condition for the scalar field is
\beq
\Phi(z_{\rm IR}) = \xi\mathbf{1}_2,
\eeq
which is stabilized by the boundary potential as the minimization of
the vacuum solution and is given by
\beq
\xi^2=\xi_0^2=\frac{m_b^2-12M_5/\sfL}{4\lambda_b},
\label{eq:scalarvac}
\eeq
the mass of the scalar is $M_\Phi^2\sfL^2=-3$, the would-be quark mass
in the model is switched off,
the indices $M,N=0,1,2,3,z$ run over all AdS$_5$, and finally $N_c$ is
the number of colors and $M_5$ is a coupling of the theory
(playing the role of $\kappa$ in the WSS model), which we have set as
$M_5\sfL=N_c/(12\pi^2)$ \cite{Bartolini:2022rkl}.

Chiral symmetry breaking is done in Ref.~\cite{Bartolini:2022rkl}
following \cite{Pomarol:2007kr,Domenech:2010aq} as
$(L_{z\mu}+R_{z\mu})_{z=z_{\rm IR}}=0$, and hence we choose
$L_z=R_z=0$ (gauge choice), $L_i=-R_i$, $\widehat{L}_0=\widehat{R}_0$
and $\Phi$ diagonal, which means that the scalar field only couples to
$L_i=-R_i$ via the covariant derivative
\beq
D_0\Phi = 0,\qquad
D_i\Phi = \p_i\Phi + 2 \i L_i\Phi,\qquad
D_z\Phi = \p_z\Phi.
\eeq
Employing the homogeneous Ansatz
\beq
L_i=-R_i=-H(z)\frac{\tau^i}{2},\qquad
\widehat{L}_0=\widehat{R}_0=\widehat{a}_0(z),\qquad
\Phi=\omega_0(z)\frac{\mathbf{1}_2}{2},
\eeq
and the coordinates $z_{\rm UV}=0$, $z_{\rm IR}=L=1$, we can write
down the vacuum solution
\beq
\Phi = \xi z^3\mathbf{1}_2,
\label{eq:Phivac}
\eeq
which holds when $H=0$ that corresponds to $\rho=0$ -- the vanishing
baryonic density and $\xi=\xi_0$ of Eq.~\eqref{eq:scalarvac}.
Once the IR potential has been fixed by choosing the two parameters
$m_b$ and $\lambda_b$ that correspond to a certain $\xi_0$, the
impact of the scalar is just to define the vacuum value of the action
in the mesonic phase of the theory.
It can readily by computed to be
\beq
S=-\lambda_b \xi_0^4.
\eeq
The boundary conditions for the fields $H(z)$ and $\widehat{a}_0(z)$
are
\begin{align}
H(0)&=0, \qquad\, H(1)=-(4\pi^2\rho)^{\frac13},\\
\widehat{a}_0(0)&= \mu,\qquad
\widehat{a}_0'(1)=0,
\end{align}
with $\mu$ being the (baryonic) chemical potential.

\begin{figure}[!htp]
  \centering
  \mbox{\subfloat[]{\includegraphics[width=0.49\linewidth]{{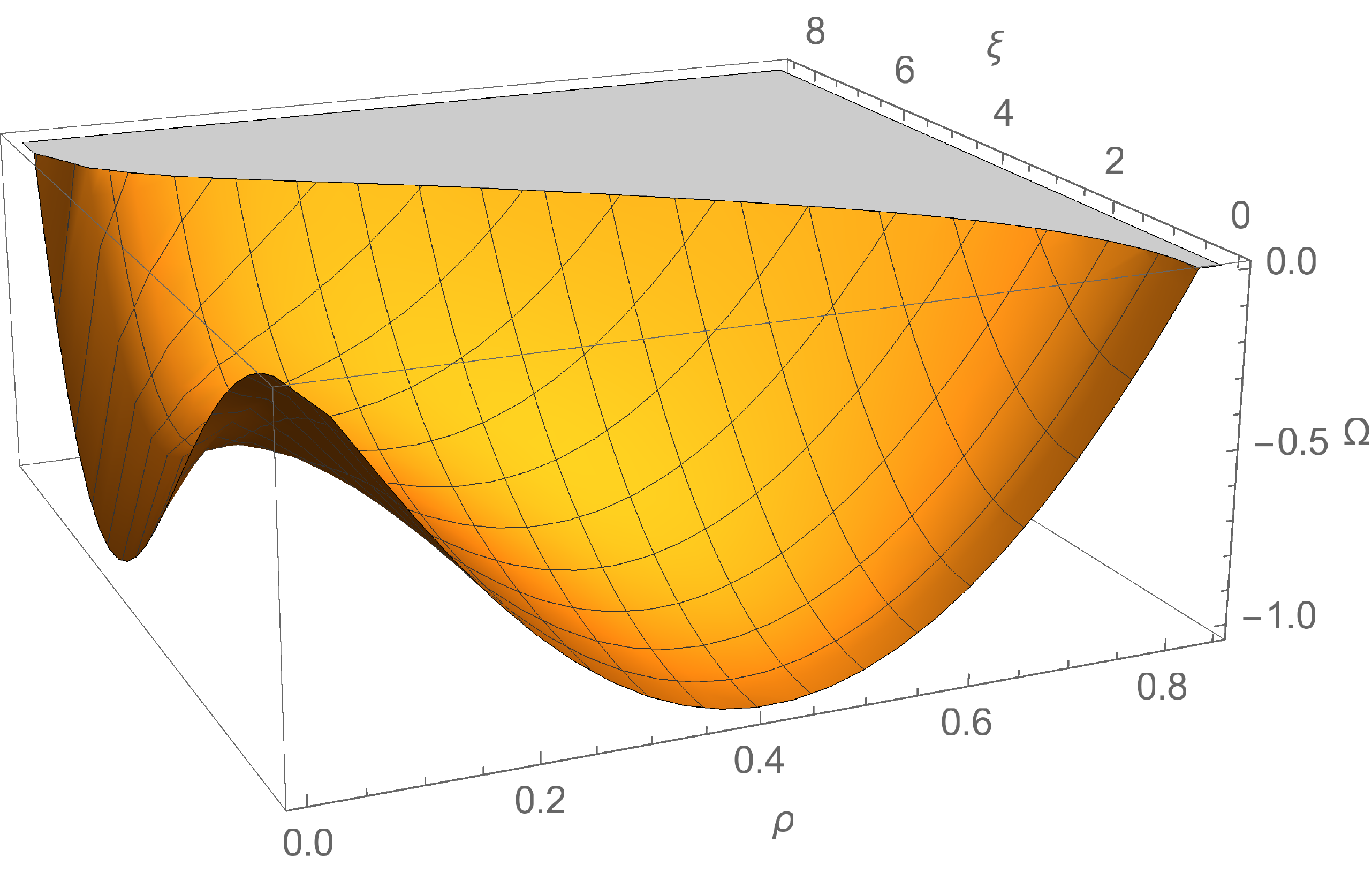}}}
    \subfloat[]{\includegraphics[width=0.49\linewidth]{{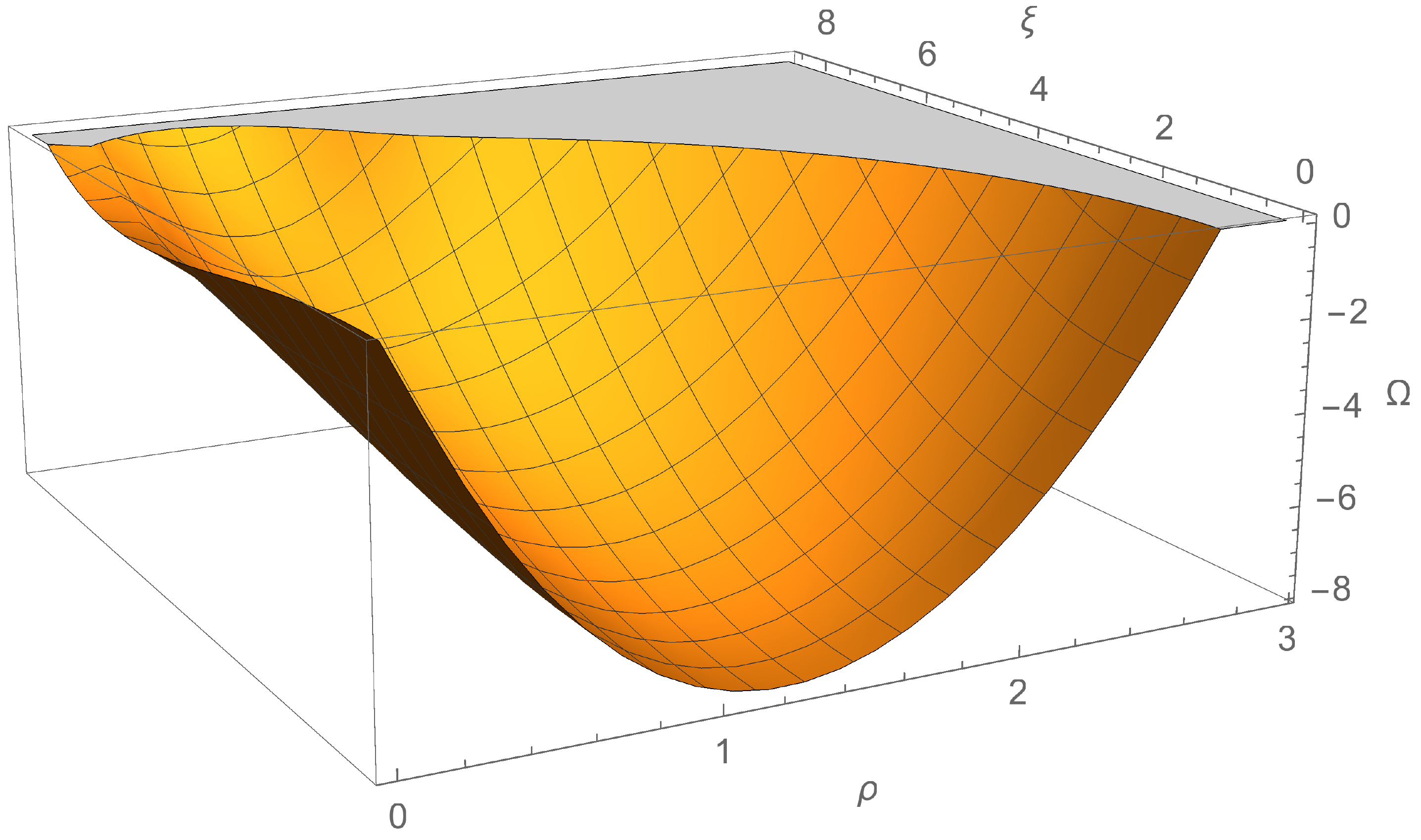}}}}
  \caption{The action evaluated as a surface for densities $\rho$ and
    scalar field with coefficient $\xi$ at (a) (for (b) twice) the
    chemical potential that corresponds to saturation density.
  In this figure, the IR potential is chosen as $m_b=0.657$ and
  $\lambda_b=0.001$, giving $\lambda_b\xi_0^4=1.024$.
  }
  \label{fig:HWvacua}
\end{figure}
In the phase $\rho>0$, the vacuum of the theory is still given by
$H=\widehat{a}_0=0$ and $\Phi$ given by the vacuum solution
\eqref{eq:Phivac} until the baryonic onset, which corresponds to the
nuclear saturation density.
At the onset, there are two vacua: a mesonic and baryonic one each
with the same value of the action (by definition), see
Fig.~\ref{fig:HWvacua}(a). 
Once $\rho>\rho_{\rm crit}$ one may ask at what configuration the
scalar field stabilizes at.
It turns out by numerical computations that the scalar turns off,
which corresponds to $\xi=0$ in the baryonic phase, see
Fig.~\ref{fig:HWvacua}(b).
This corresponds to $\Phi=0$ and when studying only the baryonic
phase, the impact of the scalar is to set the saturation density of
this simplistic hard-wall model.

\section{Comparison with large-$\lambda$ approximation}\label{app:large_lambda_approx}

In the large-$\lambda$ limit, one may consider ignoring the inclusion
of the Abelian field $L$ (in Eq.~\eqref{eq:Ansatzchi}), which however
is not a consistent choice for finite values of $\lambda$, as the
equation of motion for $L$ is not satisfied by $L=0$ (it is sourced by
$H$ and $G$, see Eq.~\eqref{eq:eomL}).

This approximation can be seen as consistent in the large-$\lambda$ limit
by considering that the field $L$ is only sourced by the Chern-Simons, which
is indeed subleading in $\lambda$ with respect to the Yang-Mills terms.
One may be led to believe that the same is true for the field $G$, and that
they have to appear at the same order in $\lambda$, but a closer look at 
its equation of motion \eqref{eq:eomG} shows that a source is present already
in the Yang-Mills terms, coming from the time derivative of the field $A_i$ 
(in our time-dependent gauge, while the same contribution arises from the 
UV boundary condition of $\widetilde{G}$ in the static gauge).

Neglecting the Abelian field $L$ was one of the approximations made in
Ref.~\cite{Kovensky:2021ddl} in addition to choosing a different
$N_c$-scaling of the isospin chemical potential (essentially defining
the large-$N_c$ baryon's proton and neutron by maximally flipping the
down and up quarks, as opposed to our definition where only one quark
is flipped from down to up).
\begin{figure}[!htp]
  \centering
  \includegraphics[width=0.5\linewidth]{{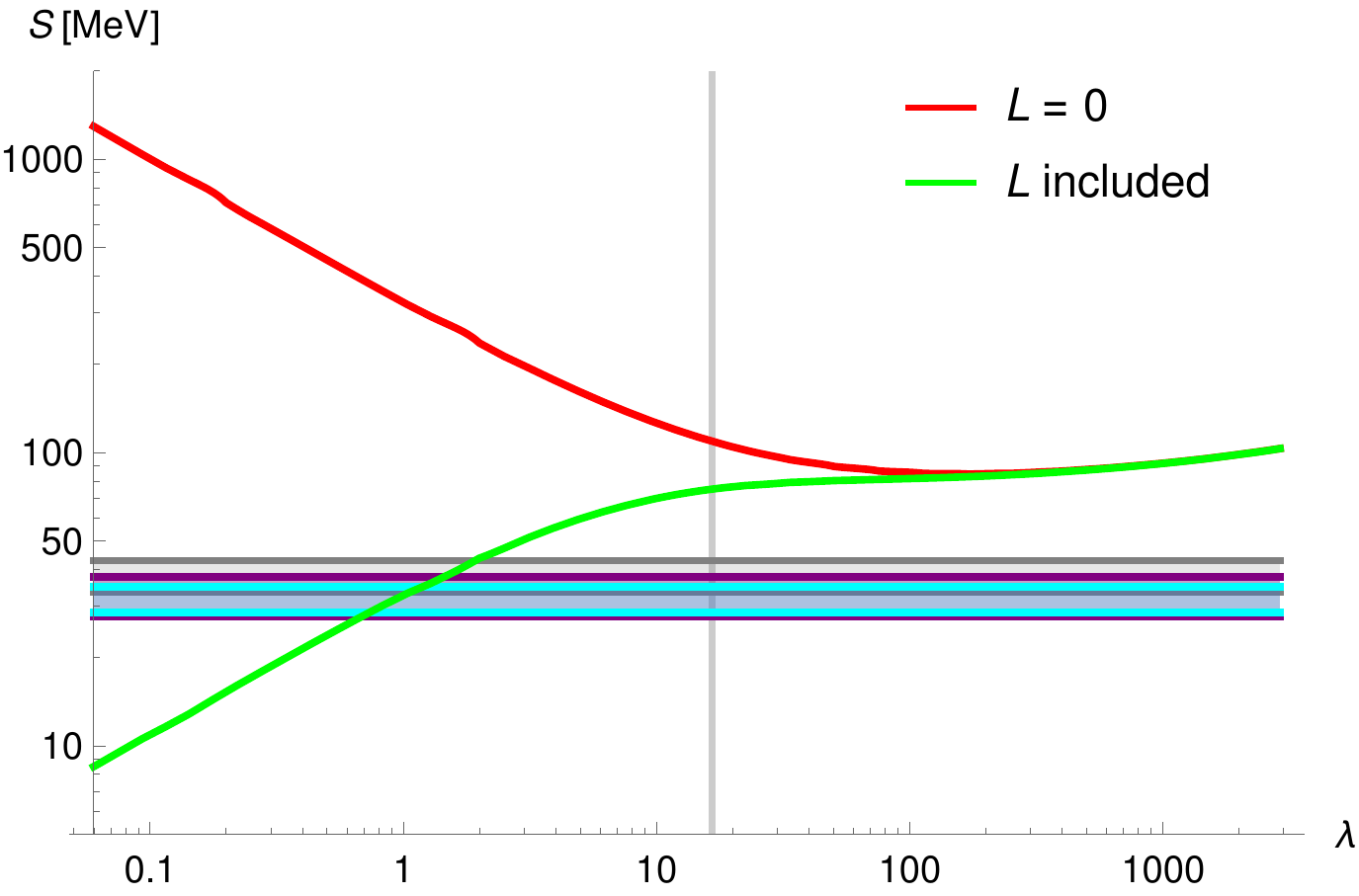}}
  \caption{
      The symmetry energy calculated as function of $\lambda$ at
      saturation density in the WSS model with $L$ taken into account
      (green solid line) and without $L$ (i.e.~$L=0$ in the Ansatz)
      (red solid line).
      In this figure, we have used $M_{\rm KK}=949\MeV$. 
  }
  \label{fig:SlambdaWrong}
\end{figure}
In order to see quantitatively how good the approximation of using
$L=0$ in the homogeneous Ansatz is, we perform the numerical calculation
corresponding to Fig.~\ref{fig:esymlambda} with and without $L$ taken
into account, see Fig.~\ref{fig:SlambdaWrong}.
From the figure, we can see that the large-$\lambda$ approximation
works in the sense that the two results asymptote to the same curve for
$\lambda\gtrsim200$.
At $\lambda=16.63$ and for $M_{\rm KK}=949\MeV$ the correct
computation including $L$ yields a symmetry energy of $74.9\MeV$
compared to $109.1\MeV$ when neglecting $L$ in the Ansatz, which at
the meson-fitted value of $\lambda$ gives an excess in the symmetry
energy of $46\%$.
The smaller $\lambda$ is the worse it gets, as expected from a
large-$\lambda$ approximation.

\bibliographystyle{JHEP}
\bibliography{bib}

\providecommand{\href}[2]{#2}\begingroup\raggedright\begin{thebibliography}{10}

\bibitem{Baldo:2016jhp}
M.~Baldo and G.~F. Burgio, {\it {The nuclear symmetry energy}},  {\em Prog.
  Part. Nucl. Phys.} {\bf 91} (2016) 203--258,
  [\href{http://arxiv.org/abs/1606.08838}{{\tt arXiv:1606.08838}}].

\bibitem{Danielewicz:2013upa}
P.~Danielewicz and J.~Lee, {\it {Symmetry Energy II: Isobaric Analog States}},
  {\em Nucl. Phys. A} {\bf 922} (2014) 1--70,
  [\href{http://arxiv.org/abs/1307.4130}{{\tt arXiv:1307.4130}}].

\bibitem{FiorellaBurgio:2018dga}
G.~Fiorella~Burgio and A.~F. Fantina, {\it {Nuclear Equation of state for
  Compact Stars and Supernovae}},  {\em Astrophys. Space Sci. Libr.} {\bf 457}
  (2018) 255--335, [\href{http://arxiv.org/abs/1804.03020}{{\tt
  arXiv:1804.03020}}].

\bibitem{Li:2021thg}
B.-A. Li, B.-J. Cai, W.-J. Xie, and N.-B. Zhang, {\it {Progress in Constraining
  Nuclear Symmetry Energy Using Neutron Star Observables Since GW170817}},
  {\em Universe} {\bf 7} (2021), no.~6 182,
  [\href{http://arxiv.org/abs/2105.04629}{{\tt arXiv:2105.04629}}].

\bibitem{Tang:2021snt}
S.-P. Tang, J.-L. Jiang, M.-Z. Han, Y.-Z. Fan, and D.-M. Wei, {\it {Constraints
  on the phase transition and nuclear symmetry parameters from PSR J0740+6620
  and multimessenger data of other neutron stars}},  {\em Phys. Rev. D} {\bf
  104} (2021), no.~6 063032, [\href{http://arxiv.org/abs/2106.04204}{{\tt
  arXiv:2106.04204}}].

\bibitem{Reed:2021nqk}
B.~T. Reed, F.~J. Fattoyev, C.~J. Horowitz, and J.~Piekarewicz, {\it
  {Implications of PREX-2 on the Equation of State of Neutron-Rich Matter}},
  {\em Phys. Rev. Lett.} {\bf 126} (2021), no.~17 172503,
  [\href{http://arxiv.org/abs/2101.03193}{{\tt arXiv:2101.03193}}].

\bibitem{Essick:2021ezp}
R.~Essick, P.~Landry, A.~Schwenk, and I.~Tews, {\it {Detailed examination of
  astrophysical constraints on the symmetry energy and the neutron skin of
  Pb208 with minimal modeling assumptions}},  {\em Phys. Rev. C} {\bf 104}
  (2021), no.~6 065804, [\href{http://arxiv.org/abs/2107.05528}{{\tt
  arXiv:2107.05528}}].

\bibitem{Gil:2021ols}
H.~Gil, P.~Papakonstantinou, and C.~H. Hyun, {\it {Constraints on the curvature
  of nuclear symmetry energy from recent astronomical data within the KIDS
  framework}},  {\em Int. J. Mod. Phys. E} {\bf 31} (2022), no.~01 2250013,
  [\href{http://arxiv.org/abs/2110.09802}{{\tt arXiv:2110.09802}}].

\bibitem{deTovar:2021sjo}
P.~B. de~Tovar, M.~Ferreira, and C.~Provid\^encia, {\it {Determination of the
  symmetry energy from the neutron star equation of state}},  {\em Phys. Rev.
  D} {\bf 104} (2021), no.~12 123036,
  [\href{http://arxiv.org/abs/2112.05551}{{\tt arXiv:2112.05551}}].

\bibitem{Barrett:2013nh}
B.~R. Barrett, P.~Navratil, and J.~P. Vary, {\it {Ab initio no core shell
  model}},  {\em Prog. Part. Nucl. Phys.} {\bf 69} (2013) 131--181.

\bibitem{Tews:2012fj}
I.~Tews, T.~Kr\"uger, K.~Hebeler, and A.~Schwenk, {\it {Neutron matter at
  next-to-next-to-next-to-leading order in chiral effective field theory}},
  {\em Phys. Rev. Lett.} {\bf 110} (2013), no.~3 032504,
  [\href{http://arxiv.org/abs/1206.0025}{{\tt arXiv:1206.0025}}].

\bibitem{Kurkela:2009gj}
A.~Kurkela, P.~Romatschke, and A.~Vuorinen, {\it {Cold Quark Matter}},  {\em
  Phys. Rev. D} {\bf 81} (2010) 105021,
  [\href{http://arxiv.org/abs/0912.1856}{{\tt arXiv:0912.1856}}].

\bibitem{Maldacena:1997re}
J.~M. Maldacena, {\it {The Large N limit of superconformal field theories and
  supergravity}},  {\em Adv. Theor. Math. Phys.} {\bf 2} (1998) 231--252,
  [\href{http://arxiv.org/abs/hep-th/9711200}{{\tt hep-th/9711200}}].

\bibitem{Witten:1998zw}
E.~Witten, {\it {Anti-de Sitter space, thermal phase transition, and
  confinement in gauge theories}},  {\em Adv. Theor. Math. Phys.} {\bf 2}
  (1998) 505--532, [\href{http://arxiv.org/abs/hep-th/9803131}{{\tt
  hep-th/9803131}}].

\bibitem{Kim:2012ey}
Y.~Kim, I.~J. Shin, and T.~Tsukioka, {\it {Holographic QCD: Past, Present, and
  Future}},  {\em Prog. Part. Nucl. Phys.} {\bf 68} (2013) 55--112,
  [\href{http://arxiv.org/abs/1205.4852}{{\tt arXiv:1205.4852}}].

\bibitem{Rebhan:2014rxa}
A.~Rebhan, {\it {The Witten-Sakai-Sugimoto model: A brief review and some
  recent results}},  {\em EPJ Web Conf.} {\bf 95} (2015) 02005,
  [\href{http://arxiv.org/abs/1410.8858}{{\tt arXiv:1410.8858}}].

\bibitem{Ammon:2015wua}
M.~Ammon and J.~Erdmenger, {\em {Gauge/gravity duality}: {Foundations and
  applications}}.
\newblock Cambridge University Press, Cambridge, 2015.

\bibitem{Sakai:2004cn}
T.~Sakai and S.~Sugimoto, {\it {Low energy hadron physics in holographic QCD}},
   {\em Prog. Theor. Phys.} {\bf 113} (2005) 843--882,
  [\href{http://arxiv.org/abs/hep-th/0412141}{{\tt hep-th/0412141}}].

\bibitem{Sakai:2005yt}
T.~Sakai and S.~Sugimoto, {\it {More on a holographic dual of QCD}},  {\em
  Prog. Theor. Phys.} {\bf 114} (2005) 1083--1118,
  [\href{http://arxiv.org/abs/hep-th/0507073}{{\tt hep-th/0507073}}].

\bibitem{Gursoy:2007cb}
U.~Gursoy and E.~Kiritsis, {\it {Exploring improved holographic theories for
  QCD: Part I}},  {\em JHEP} {\bf 02} (2008) 032,
  [\href{http://arxiv.org/abs/0707.1324}{{\tt arXiv:0707.1324}}].

\bibitem{Gursoy:2007er}
U.~Gursoy, E.~Kiritsis, and F.~Nitti, {\it {Exploring improved holographic
  theories for QCD: Part II}},  {\em JHEP} {\bf 02} (2008) 019,
  [\href{http://arxiv.org/abs/0707.1349}{{\tt arXiv:0707.1349}}].

\bibitem{Jarvinen:2011qe}
M.~Jarvinen and E.~Kiritsis, {\it {Holographic Models for QCD in the Veneziano
  Limit}},  {\em JHEP} {\bf 03} (2012) 002,
  [\href{http://arxiv.org/abs/1112.1261}{{\tt arXiv:1112.1261}}].

\bibitem{Alho:2013hsa}
T.~Alho, M.~J\"arvinen, K.~Kajantie, E.~Kiritsis, C.~Rosen, and K.~Tuominen,
  {\it {A holographic model for QCD in the Veneziano limit at finite
  temperature and density}},  {\em JHEP} {\bf 04} (2014) 124,
  [\href{http://arxiv.org/abs/1312.5199}{{\tt arXiv:1312.5199}}]. [Erratum:
  JHEP 02, 033 (2015)].

\bibitem{Jarvinen:2021jbd}
M.~J\"arvinen, {\it {Holographic modeling of nuclear matter and neutron
  stars}},  {\em Eur. Phys. J. C} {\bf 82} (2022), no.~4 282,
  [\href{http://arxiv.org/abs/2110.08281}{{\tt arXiv:2110.08281}}].

\bibitem{Polchinski:2001tt}
J.~Polchinski and M.~J. Strassler, {\it {Hard scattering and gauge/string
  duality}},  {\em Phys. Rev. Lett.} {\bf 88} (2002) 031601,
  [\href{http://arxiv.org/abs/hep-th/0109174}{{\tt hep-th/0109174}}].

\bibitem{Boschi-Filho:2002wdj}
H.~Boschi-Filho and N.~R.~F. Braga, {\it {QCD/string holographic mapping and
  glueball mass spectrum}},  {\em Eur. Phys. J. C} {\bf 32} (2004) 529--533,
  [\href{http://arxiv.org/abs/hep-th/0209080}{{\tt hep-th/0209080}}].

\bibitem{Polchinski:2002jw}
J.~Polchinski and M.~J. Strassler, {\it {Deep inelastic scattering and
  gauge/string duality}},  {\em JHEP} {\bf 05} (2003) 012,
  [\href{http://arxiv.org/abs/hep-th/0209211}{{\tt hep-th/0209211}}].

\bibitem{Boschi-Filho:2002xih}
H.~Boschi-Filho and N.~R.~F. Braga, {\it {Gauge/string duality and scalar
  glueball mass ratios}},  {\em JHEP} {\bf 05} (2003) 009,
  [\href{http://arxiv.org/abs/hep-th/0212207}{{\tt hep-th/0212207}}].

\bibitem{deTeramond:2005su}
G.~F. de~Teramond and S.~J. Brodsky, {\it {Hadronic spectrum of a holographic
  dual of QCD}},  {\em Phys. Rev. Lett.} {\bf 94} (2005) 201601,
  [\href{http://arxiv.org/abs/hep-th/0501022}{{\tt hep-th/0501022}}].

\bibitem{Erlich:2005qh}
J.~Erlich, E.~Katz, D.~T. Son, and M.~A. Stephanov, {\it {QCD and a holographic
  model of hadrons}},  {\em Phys. Rev. Lett.} {\bf 95} (2005) 261602,
  [\href{http://arxiv.org/abs/hep-ph/0501128}{{\tt hep-ph/0501128}}].

\bibitem{DaRold:2005vr}
L.~Da~Rold and A.~Pomarol, {\it {The Scalar and pseudoscalar sector in a
  five-dimensional approach to chiral symmetry breaking}},  {\em JHEP} {\bf 01}
  (2006) 157, [\href{http://arxiv.org/abs/hep-ph/0510268}{{\tt
  hep-ph/0510268}}].

\bibitem{Hirn:2005nr}
J.~Hirn and V.~Sanz, {\it {Interpolating between low and high energy QCD via a
  5-D Yang-Mills model}},  {\em JHEP} {\bf 12} (2005) 030,
  [\href{http://arxiv.org/abs/hep-ph/0507049}{{\tt hep-ph/0507049}}].

\bibitem{Karch:2006pv}
A.~Karch, E.~Katz, D.~T. Son, and M.~A. Stephanov, {\it {Linear confinement and
  AdS/QCD}},  {\em Phys. Rev. D} {\bf 74} (2006) 015005,
  [\href{http://arxiv.org/abs/hep-ph/0602229}{{\tt hep-ph/0602229}}].

\bibitem{Kim:2010dp}
Y.~Kim, Y.~Seo, I.~J. Shin, and S.-J. Sin, {\it {Symmetry energy of dense
  matter in holographic QCD}},  {\em JHEP} {\bf 06} (2011) 011,
  [\href{http://arxiv.org/abs/1011.0868}{{\tt arXiv:1011.0868}}].

\bibitem{Kovensky:2021ddl}
N.~Kovensky and A.~Schmitt, {\it {Isospin asymmetry in holographic baryonic
  matter}},  {\em SciPost Phys.} {\bf 11} (2021), no.~2 029,
  [\href{http://arxiv.org/abs/2105.03218}{{\tt arXiv:2105.03218}}].

\bibitem{Kovensky:2021kzl}
N.~Kovensky, A.~Poole, and A.~Schmitt, {\it {Building a realistic neutron star
  from holography}},  {\em Phys. Rev. D} {\bf 105} (2022), no.~3 034022,
  [\href{http://arxiv.org/abs/2111.03374}{{\tt arXiv:2111.03374}}].

\bibitem{Bartolini:2022rkl}
L.~Bartolini, S.~B. Gudnason, J.~Leutgeb, and A.~Rebhan, {\it {Neutron stars
  and phase diagram in a hard-wall AdS/QCD model}},  {\em Phys. Rev. D} {\bf
  105} (2022), no.~12 126014, [\href{http://arxiv.org/abs/2202.12845}{{\tt
  arXiv:2202.12845}}].

\bibitem{Hata:2007mb}
H.~Hata, T.~Sakai, S.~Sugimoto, and S.~Yamato, {\it {Baryons from instantons in
  holographic QCD}},  {\em Prog. Theor. Phys.} {\bf 117} (2007) 1157,
  [\href{http://arxiv.org/abs/hep-th/0701280}{{\tt hep-th/0701280}}].

\bibitem{Hashimoto:2008zw}
K.~Hashimoto, T.~Sakai, and S.~Sugimoto, {\it {Holographic Baryons: Static
  Properties and Form Factors from Gauge/String Duality}},  {\em Prog. Theor.
  Phys.} {\bf 120} (2008) 1093--1137,
  [\href{http://arxiv.org/abs/0806.3122}{{\tt arXiv:0806.3122}}].

\bibitem{Bolognesi:2013nja}
S.~Bolognesi and P.~Sutcliffe, {\it {The Sakai-Sugimoto soliton}},  {\em JHEP}
  {\bf 01} (2014) 078, [\href{http://arxiv.org/abs/1309.1396}{{\tt
  arXiv:1309.1396}}].

\bibitem{Erdmenger:2007ap}
J.~Erdmenger, M.~Kaminski, and F.~Rust, {\it {Isospin diffusion in thermal
  AdS/CFT with flavor}},  {\em Phys. Rev. D} {\bf 76} (2007) 046001,
  [\href{http://arxiv.org/abs/0704.1290}{{\tt arXiv:0704.1290}}].

\bibitem{Erdmenger:2007ja}
J.~Erdmenger, M.~Kaminski, and F.~Rust, {\it {Holographic vector mesons from
  spectral functions at finite baryon or isospin density}},  {\em Phys. Rev. D}
  {\bf 77} (2008) 046005, [\href{http://arxiv.org/abs/0710.0334}{{\tt
  arXiv:0710.0334}}].

\bibitem{Erdmenger:2008yj}
J.~Erdmenger, M.~Kaminski, P.~Kerner, and F.~Rust, {\it {Finite baryon and
  isospin chemical potential in AdS/CFT with flavor}},  {\em JHEP} {\bf 11}
  (2008) 031, [\href{http://arxiv.org/abs/0807.2663}{{\tt arXiv:0807.2663}}].

\bibitem{Parnachev:2007bc}
A.~Parnachev, {\it {Holographic QCD with Isospin Chemical Potential}},  {\em
  JHEP} {\bf 02} (2008) 062, [\href{http://arxiv.org/abs/0708.3170}{{\tt
  arXiv:0708.3170}}].

\bibitem{Aharony:2007uu}
O.~Aharony, K.~Peeters, J.~Sonnenschein, and M.~Zamaklar, {\it {Rho meson
  condensation at finite isospin chemical potential in a holographic model for
  QCD}},  {\em JHEP} {\bf 02} (2008) 071,
  [\href{http://arxiv.org/abs/0709.3948}{{\tt arXiv:0709.3948}}].

\bibitem{Adkins:1983ya}
G.~S. Adkins, C.~R. Nappi, and E.~Witten, {\it {Static Properties of Nucleons
  in the Skyrme Model}},  {\em Nucl. Phys. B} {\bf 228} (1983) 552.

\bibitem{Lee:2010sw}
H.~K. Lee, B.-Y. Park, and M.~Rho, {\it {Half-Skyrmions, Tensor Forces and
  Symmetry Energy in Cold Dense Matter}},  {\em Phys. Rev. C} {\bf 83} (2011)
  025206, [\href{http://arxiv.org/abs/1005.0255}{{\tt arXiv:1005.0255}}].
  [Erratum: Phys.Rev.C 84, 059902 (2011)].

\bibitem{Adam:2022aes}
C.~Adam, A.~Garcia Martin-Caro, M.~Huidobro, R.~V\'azquez, and
  A.~Wereszczynski, {\it {Quantum skyrmion crystals and the symmetry energy of
  dense matter}},  {\em Phys. Rev. D} {\bf 106} (2022), no.~11 114031,
  [\href{http://arxiv.org/abs/2202.00953}{{\tt arXiv:2202.00953}}].

\bibitem{Rozali:2007rx}
M.~Rozali, H.-H. Shieh, M.~Van~Raamsdonk, and J.~Wu, {\it {Cold Nuclear Matter
  In Holographic QCD}},  {\em JHEP} {\bf 01} (2008) 053,
  [\href{http://arxiv.org/abs/0708.1322}{{\tt arXiv:0708.1322}}].

\bibitem{Elliot-Ripley:2016uwb}
M.~Elliot-Ripley, P.~Sutcliffe, and M.~Zamaklar, {\it {Phases of kinky
  holographic nuclear matter}},  {\em JHEP} {\bf 10} (2016) 088,
  [\href{http://arxiv.org/abs/1607.04832}{{\tt arXiv:1607.04832}}].

\bibitem{Li:2015uea}
S.-w. Li, A.~Schmitt, and Q.~Wang, {\it {From holography towards real-world
  nuclear matter}},  {\em Phys. Rev. D} {\bf 92} (2015), no.~2 026006,
  [\href{http://arxiv.org/abs/1505.04886}{{\tt arXiv:1505.04886}}].

\bibitem{Bigazzi:2018cpg}
F.~Bigazzi and P.~Niro, {\it {Neutron-proton mass difference from gauge/gravity
  duality}},  {\em Phys. Rev. D} {\bf 98} (2018), no.~4 046004,
  [\href{http://arxiv.org/abs/1803.05202}{{\tt arXiv:1803.05202}}].

\bibitem{Kovensky:2019bih}
N.~Kovensky and A.~Schmitt, {\it {Heavy Holographic QCD}},  {\em JHEP} {\bf 02}
  (2020) 096, [\href{http://arxiv.org/abs/1911.08433}{{\tt arXiv:1911.08433}}].

\bibitem{Kovensky:2020xif}
N.~Kovensky and A.~Schmitt, {\it {Holographic quarkyonic matter}},  {\em JHEP}
  {\bf 09} (2020) 112, [\href{http://arxiv.org/abs/2006.13739}{{\tt
  arXiv:2006.13739}}].

\bibitem{Kovensky:2023mye}
N.~Kovensky, A.~Poole, and A.~Schmitt, {\it {Phases of cold holographic QCD:
  Baryons, pions and rho mesons}},  {\em SciPost Phys.} {\bf 15} (2023), no.~4
  162, [\href{http://arxiv.org/abs/2302.10675}{{\tt arXiv:2302.10675}}].

\bibitem{Singh:2022obu}
A.~Singh and K.~P. Yogendran, {\it {Phases of a 10-D holographic hard wall
  model}},  {\em JHEP} {\bf 02} (2023) 168,
  [\href{http://arxiv.org/abs/2208.09387}{{\tt arXiv:2208.09387}}].

\bibitem{CruzRojas:2023ugm}
J.~Cruz~Rojas, T.~Demircik, and M.~J\"arvinen, {\it {Popcorn Transitions and
  Approach to Conformality in Homogeneous Holographic Nuclear Matter}},  {\em
  Symmetry} {\bf 15} (2023), no.~2 331,
  [\href{http://arxiv.org/abs/2301.03173}{{\tt arXiv:2301.03173}}].

\bibitem{Bartolini:2023eam}
L.~Bartolini and S.~B. Gudnason, {\it {Boundary terms in the
  Witten-Sakai-Sugimoto model at finite density}},  {\em Phys. Rev. D} {\bf
  109} (2024), no.~6 066006, [\href{http://arxiv.org/abs/2309.16328}{{\tt
  arXiv:2309.16328}}].

\bibitem{Ilderton:2005yu}
A.~Ilderton, {\it {Radial evolution in anti-de Sitter spacetime}},  {\em Int.
  J. Mod. Phys. A} {\bf 21} (2006) 3289--3294,
  [\href{http://arxiv.org/abs/hep-th/0501218}{{\tt hep-th/0501218}}].

\bibitem{Baldino:2017mqq}
S.~Baldino, S.~Bolognesi, S.~B. Gudnason, and D.~Koksal, {\it {Solitonic
  approach to holographic nuclear physics}},  {\em Phys. Rev. D} {\bf 96}
  (2017), no.~3 034008, [\href{http://arxiv.org/abs/1703.08695}{{\tt
  arXiv:1703.08695}}].

\bibitem{Baldino:2021uie}
S.~Baldino, L.~Bartolini, S.~Bolognesi, and S.~B. Gudnason, {\it {Holographic
  Nuclear Physics with Massive Quarks}},  {\em Phys. Rev. D} {\bf 103} (2021),
  no.~12 126015, [\href{http://arxiv.org/abs/2102.00680}{{\tt
  arXiv:2102.00680}}].

\bibitem{Bartolini:2023wis}
L.~Bartolini and S.~B. Gudnason, {\it {Neutron stars in the
  Witten-Sakai-Sugimoto model}},  {\em JHEP} {\bf 11} (2023) 209,
  [\href{http://arxiv.org/abs/2307.11886}{{\tt arXiv:2307.11886}}].

\bibitem{Lim:2017aac}
Y.~Lim and J.~W. Holt, {\it {Proton pairing in neutron stars from chiral
  effective field theory}},  {\em Phys. Rev. C} {\bf 103} (2021), no.~2 025807,
  [\href{http://arxiv.org/abs/1709.08793}{{\tt arXiv:1709.08793}}].

\bibitem{Park:2009mg}
B.-Y. Park, J.-I. Kim, and M.~Rho, {\it {Kaons in Dense Half-Skyrmion Matter}},
   {\em Phys. Rev. C} {\bf 81} (2010) 035203,
  [\href{http://arxiv.org/abs/0912.3213}{{\tt arXiv:0912.3213}}].

\bibitem{Rho:2009ym}
M.~Rho, S.-J. Sin, and I.~Zahed, {\it {Dense QCD: A Holographic Dyonic Salt}},
  {\em Phys. Lett. B} {\bf 689} (2010) 23--27,
  [\href{http://arxiv.org/abs/0910.3774}{{\tt arXiv:0910.3774}}].

\bibitem{Kaplunovsky:2012gb}
V.~Kaplunovsky, D.~Melnikov, and J.~Sonnenschein, {\it {Baryonic Popcorn}},
  {\em JHEP} {\bf 11} (2012) 047, [\href{http://arxiv.org/abs/1201.1331}{{\tt
  arXiv:1201.1331}}].

\bibitem{Adam:2023cee}
C.~Adam, A.~Garcia Martin-Caro, M.~Huidobro, and A.~Wereszczynski, {\it {Skyrme
  Crystals, Nuclear Matter and Compact Stars}},  {\em Symmetry} {\bf 15}
  (2023), no.~4 899, [\href{http://arxiv.org/abs/2305.06639}{{\tt
  arXiv:2305.06639}}].

\bibitem{Pomarol:2007kr}
A.~Pomarol and A.~Wulzer, {\it {Stable skyrmions from extra dimensions}},  {\em
  JHEP} {\bf 03} (2008) 051--051, [\href{http://arxiv.org/abs/0712.3276}{{\tt
  arXiv:0712.3276}}].

\bibitem{Domenech:2010aq}
O.~Domenech, G.~Panico, and A.~Wulzer, {\it {Massive Pions, Anomalies and
  Baryons in Holographic QCD}},  {\em Nucl. Phys. A} {\bf 853} (2011) 97--123,
  [\href{http://arxiv.org/abs/1009.0711}{{\tt arXiv:1009.0711}}].

\end{thebibliography}\endgroup

\end{document}